\documentclass[longauth]{aa}  
\usepackage{graphicx}
\usepackage[dvipsnames]{xcolor}
\usepackage{caption}
\usepackage{subcaption}
\usepackage{arydshln}
\usepackage{txfonts}
\usepackage{orcidlink}
\usepackage{verbatim}
\usepackage{multirow}
\usepackage{multicol}
\usepackage{float}
\usepackage{upgreek}
\usepackage{placeins}
\usepackage{orcidlink}
\usepackage{hyperref}
\hypersetup{colorlinks=true, linkcolor=blue, urlcolor=blue, filecolor=blue, citecolor=blue}
\usepackage[normalem]{ulem}

\usepackage{soul}
\let\ACMmaketitle=\maketitle
\renewcommand{\maketitle}{\begingroup\let\footnote=\thanks \ACMmaketitle\endgroup}
\newcommand{\mjup}[0]{\,M_\mathrm{Jup}}
\newcommand{\rjup}[0]{\,R_\mathrm{Jup}}
\newcommand{\rearth}[0]{\,R_\oplus}

\begin{document} 
   \title{TOI-3288 b and TOI-4666 b: two gas giants transiting low-mass stars characterised by NIRPS}

\author{
Yolanda G. C. Frensch\inst{\ref{geneva},\ref{esosantiago},*}\orcidlink{0000-0003-4009-0330},
Fran\c{c}ois Bouchy\inst{\ref{geneva}}\orcidlink{0000-0002-7613-393X},
Gaspare Lo Curto\inst{\ref{esosantiago}}\orcidlink{0000-0002-1158-9354},
Alexandrine L'Heureux\inst{\ref{irex}}\orcidlink{0009-0005-6135-6769},
Roseane de Lima Gomes\inst{\ref{irex},\ref{ufrn}}\orcidlink{0000-0002-2023-7641},
Jo\~ao Faria\inst{\ref{geneva},\ref{caup}},
Xavier Dumusque\inst{\ref{geneva}}\orcidlink{0000-0002-9332-2011},
Lison Malo\inst{\ref{irex},\ref{omm}}\orcidlink{0000-0002-8786-8499},
Marion Cointepas\inst{\ref{geneva},\ref{ipag}}\orcidlink{0009-0001-6168-2178},
Avidaan Srivastava\inst{\ref{irex}}\orcidlink{0009-0009-7136-1528},
Xavier Bonfils\inst{\ref{ipag}}\orcidlink{0000-0001-9003-8894},
Elisa Delgado-Mena\inst{\ref{cab},\ref{caup}}\orcidlink{0000-0003-4434-2195},
Nicola Nari\inst{\ref{lb},\ref{iac},\ref{depastrolaguna}},
\'Etienne Artigau\inst{\ref{irex},\ref{omm}}\orcidlink{0000-0003-3506-5667},
Fr\'ed\'erique Baron\inst{\ref{irex},\ref{omm}}\orcidlink{0000-0002-5074-1128},
Susana C. C. Barros\inst{\ref{caup},\ref{depfisporto}}\orcidlink{0000-0003-2434-3625},
Bj\"orn Benneke\inst{\ref{ucla},\ref{irex}}\orcidlink{0000-0001-5578-1498},
Marta Bryan\inst{\ref{toronto}},
Bruno L. Canto Martins\inst{\ref{ufrn}}\orcidlink{0000-0001-5578-7400},
Izan de Castro Le\~ao\inst{\ref{ufrn}}\orcidlink{0000-0001-5845-947X},
Ryan Cloutier\inst{\ref{mcmaster}}\orcidlink{0000-0001-5383-9393},
Neil J. Cook\inst{\ref{irex}}\orcidlink{0000-0003-4166-4121},
Nicolas B. Cowan\inst{\ref{mcgill},\ref{mcgill_planetary}}\orcidlink{0000-0001-6129-5699},
Eduardo Cristo\inst{\ref{caup}},
Jose R. De Medeiros\inst{\ref{ufrn}}\orcidlink{0000-0001-8218-1586},
Xavier Delfosse\inst{\ref{ipag}}\orcidlink{0000-0001-5099-7978},
Ren\'e Doyon\inst{\ref{irex},\ref{omm}}\orcidlink{0000-0001-5485-4675},
David Ehrenreich\inst{\ref{geneva},\ref{cvu}},
Jonay I. Gonz\'alez Hern\'andez\inst{\ref{iac},\ref{depastrolaguna}}\orcidlink{0000-0002-0264-7356},
David Lafreni\`ere\inst{\ref{irex}}\orcidlink{0000-0002-6780-4252},
Christophe Lovis\inst{\ref{geneva}}\orcidlink{0000-0001-7120-5837},
Claudio Melo\inst{\ref{esogarching}},
Lucile Mignon\inst{\ref{geneva},\ref{ipag}},
Christoph Mordasini\inst{\ref{bern}}\orcidlink{0000-0002-1013-2811},
Francesco Pepe\inst{\ref{geneva}}\orcidlink{0000-0002-9815-773X},
Rafael Rebolo\inst{\ref{iac},\ref{depastrolaguna},\ref{csic}}\orcidlink{0000-0003-3767-7085},
Jason Rowe\inst{\ref{bishop}},
Nuno C. Santos\inst{\ref{caup},\ref{depfisporto}}\orcidlink{0000-0003-4422-2919},
Damien S\'egransan\inst{\ref{geneva}},
Alejandro Su\'arez Mascare\~no\inst{\ref{iac},\ref{depastrolaguna}}\orcidlink{0000-0002-3814-5323},
St\'ephane Udry\inst{\ref{geneva}}\orcidlink{0000-0001-7576-6236},
Diana Valencia\inst{\ref{toronto}}\orcidlink{0000-0003-3993-4030},
Gregg Wade\inst{\ref{queen},\ref{rmc}},
Khaled Al Moulla\inst{\ref{caup},\ref{geneva}}\orcidlink{0000-0002-3212-5778},
Romain Allart\inst{\ref{irex}}\orcidlink{0000-0002-1199-9759},
Jose M. Almenara\inst{\ref{ipag}}\orcidlink{0000-0003-3208-9815},
Khalid Barkaoui\inst{\ref{iac},\ref{aru_liege},\ref{mit_eaps}}\orcidlink{0000-0003-1464-9276},
Charles Cadieux\inst{\ref{irex}}\orcidlink{0000-0001-9291-5555},
Amadeo Castro-Gonz\'alez\inst{\ref{geneva}}\orcidlink{0000-0001-7439-3618},
Karen A. Collins\inst{\ref{cfa}}\orcidlink{0000-0001-6588-9574},
Sergio B. Fajardo-Acosta\inst{\ref{caltech_ipac}},
Thierry Forveille\inst{\ref{ipag}}\orcidlink{0000-0003-0536-4607},
Tianjun Gan\inst{\ref{hangzhou}}\orcidlink{0000-0002-4503-9705},
Jo\~ao Gomes da Silva\inst{\ref{caup}}\orcidlink{0000-0001-8056-9202},
Nolan Grieves\inst{\ref{geneva}}\orcidlink{0000-0001-8105-0373},
Melissa J. Hobson\inst{\ref{geneva}}\orcidlink{0000-0002-5945-7975},
Steve Howell\inst{\ref{ames}}\orcidlink{0000-0002-2532-2853},
Pierrot Lamontagne\inst{\ref{irex}},
Lina Messamah\inst{\ref{geneva}}\orcidlink{0000-0003-0029-2835},
Louise D. Nielsen\inst{\ref{geneva},\ref{esogarching},\ref{lmu}}\orcidlink{0000-0002-5254-2499},
Ares Osborn\inst{\ref{mcmaster},\ref{ipag}}\orcidlink{0000-0002-5899-7750},
L\'ena Parc\inst{\ref{geneva}}\orcidlink{0000-0002-7382-1913},
Caroline Piaulet-Ghorayeb\inst{\ref{irex},\ref{chicago}}\orcidlink{0000-0002-2875-917X},
Keivan G. Stassun\inst{\ref{vanderbilt}}\orcidlink{0000-0002-3481-9052},
Atanas K. Stefanov\inst{\ref{iac},\ref{depastrolaguna}}\orcidlink{0000-0002-6059-1178},
Stephanie Striegel\inst{\ref{seti}}\orcidlink{0009-0008-5145-0446},
Sol\`ene Ulmer-Moll\inst{\ref{geneva}}\orcidlink{0000-0003-2417-7006},
Valentina Vaulato\inst{\ref{geneva}}\orcidlink{0000-0001-7329-3471},
Cristilyn N. Watkins\inst{\ref{cfa}}\orcidlink{0000-0001-8621-6731}
}

\institute{ 
Observatoire de Gen\`eve, D\'epartement d’Astronomie, Universit\'e de Gen\`eve, Chemin Pegasi 51, 1290 Versoix, Switzerland\label{geneva} 
\\  \inst{*}\email{yolanda.frensch@unige.ch}
\and
European Southern Observatory (ESO), Av. Alonso de Cordova 3107,  Casilla 19001, Santiago de Chile, Chile\label{esosantiago} 
\and
Institut Trottier de recherche sur les exoplan\`etes, D\'epartement de Physique, Universit\'e de Montr\'eal, Montr\'eal, Qu\'ebec, Canada\label{irex} 
\and
Departamento de F\'isica Te\'orica e Experimental, Universidade Federal do Rio Grande do Norte, Campus Universit\'ario, Natal, RN, 59072-970, Brazil\label{ufrn} 
\and
Instituto de Astrof\'isica e Ci\^encias do Espa\c{c}o, Universidade do Porto, CAUP, Rua das Estrelas, 4150-762 Porto, Portugal\label{caup} 
\and
Observatoire du Mont-M\'egantic, Qu\'ebec, Canada\label{omm} 
\and
Univ. Grenoble Alpes, CNRS, IPAG, F-38000 Grenoble, France\label{ipag} 
\and
Centro de Astrobiolog\'ia (CAB), CSIC-INTA, Camino Bajo del Castillo s/n, 28692, Villanueva de la Ca\~nada (Madrid), Spain\label{cab} 
\and
Light Bridges S.L., Observatorio del Teide, Carretera del Observatorio, s/n Guimar, 38500, Tenerife, Canarias, Spain\label{lb} 
\and
Instituto de Astrof\'isica de Canarias (IAC), Calle V\'ia L\'actea s/n, 38205 La Laguna, Tenerife, Spain\label{iac} 
\and
Departamento de Astrof\'isica, Universidad de La Laguna (ULL), 38206 La Laguna, Tenerife, Spain\label{depastrolaguna} 
\and 
Departamento de F\'isica e Astronomia, Faculdade de Ci\^encias, Universidade do Porto, Rua do Campo Alegre, 4169-007 Porto, Portugal\label{depfisporto} 
\and
Department of Earth, Planetary, and Space Sciences, University of California, Los Angeles, CA 90095, USA\label{ucla} 
\and
Department of Physics, University of Toronto, Toronto, ON M5S 3H4, Canada\label{toronto} 
\and
Department of Physics \& Astronomy, McMaster University, 1280 Main St W, Hamilton, ON, L8S 4L8, Canada\label{mcmaster} 
\and
Department of Physics, McGill University, 3600 rue University, Montr\'eal, QC, H3A 2T8, Canada\label{mcgill} 
\and
Department of Earth \& Planetary Sciences, McGill University, 3450 rue University, Montr\'eal, QC, H3A 0E8, Canada\label{mcgill_planetary} 
\and 
Centre Vie dans l’Univers, Facult\'e des sciences de l’Universit\'e de Gen\`eve, Quai Ernest-Ansermet 30, 1205 Geneva, Switzerland\label{cvu} 
\and
European Southern Observatory (ESO), Karl-Schwarzschild-Str. 2, 85748 Garching bei M\"unchen, Germany\label{esogarching} 
\and
Space Research and Planetary Sciences, Physics Institute, University of Bern, Gesellschaftsstrasse 6, 3012 Bern, Switzerland\label{bern} 
\and
Consejo Superior de Investigaciones Cient\'ificas (CSIC), E-28006 Madrid, Spain\label{csic} 
\and
Bishop's University, Dept of Physics and Astronomy, Johnson-104E, 2600 College Street, Sherbrooke, QC, Canada, J1M 1Z7, Canada\label{bishop} 
\and
Department of Physics, Engineering Physics, and Astronomy, Queen’s University, 99 University Avenue, Kingston, ON K7L 3N6, Canada\label{queen} 
\and
Department of Physics and Space Science, Royal Military College of Canada, 13 General Crerar Cres., Kingston, ON K7P 2M3, Canada\label{rmc} 
\and
Astrobiology Research Unit, Universit\'e de Li\`ege, 19C All\'ee du 6 Ao\^ut, 4000 Li\`ege, Belgium\label{aru_liege} 
\and
Department of Earth, Atmospheric and Planetary Sciences, Massachusetts Institute of Technology, Cambridge, MA 02139, USA\label{mit_eaps} 
\and
Center for astrophysics $\vert$ Harvard \& Smithsonian, 60 Garden Street, Cambridge, MA 02138, USA\label{cfa} 
\and
Caltech/IPAC, Mail Code 100-22, Pasadena, CA 91125, USA\label{caltech_ipac} 
\and
Department of Astronomy, Westlake University,Hangzhou 310030, Zhejiang Province, People's Republic of China\label{hangzhou} 
\and
NASA Ames Research Center, Moffett Field, CA 94035, USA\label{ames} 
\and
University Observatory, Faculty of Physics, Ludwig-Maximilians-Universit\"at M\"unchen, Scheinerstr. 1, 81679 Munich, Germany\label{lmu} 
\and
Department of Astronomy \& Astrophysics, University of Chicago, 5640 South Ellis Avenue, Chicago, IL 60637, USA\label{chicago} 
\and
Department of Physics and Astronomy, Vanderbilt University, VU Station 1807, Nashville, TN 37235, USA\label{vanderbilt} 
\and
SETI Institute, Mountain View, CA 94043, USA NASA Ames Research Center, Moffett Field, CA 94035, USA\label{seti} 
}
    \authorrunning{Y.G.C. Frensch et al.}
    \titlerunning{Two gas giants transiting low-mass stars revealed by NIRPS}
   \date{Received 10 October 2025; accepted 20 December 2025}
 
  \abstract
  {Gas giant planets orbiting low-mass stars ($T_\mathrm{eff} \lesssim 4600$ K) are uncommon outcomes of planet formation. Increasing the sample of well-characterised giants around early M dwarfs will enable population-level studies of their properties, offering valuable insights into their formation and evolutionary histories.}
   {We aim to confirm and characterise giant exoplanets transiting M dwarfs identified by the \textit{TESS} mission. To this end, we have started the Gas giAnts Transiting lOw-mass Stars (GATOS) program within the NIRPS Guaranteed Time Observations (GTO).}
   {High-resolution spectroscopic data are obtained in the optical and near-infrared (nIR), combining HARPS and NIRPS. We derive radial velocities (RVs) via the cross-correlation function and implement a novel post-processing procedure to further mitigate telluric contamination in the nIR. The resulting RVs are jointly fit with \textit{TESS} and ground-based photometry to derive the orbital and physical parameters of the systems.}
   {We present the GATOS program and its first results. We confirm two gas giants transiting the low-mass stars TOI-3288~A (K9V, $T_{\rm eff} = 3933 \pm 48$~K) and TOI-4666 (M2.5V, $T_{\rm eff} = 3512\pm 36$~K). TOI-3288~A hosts a Hot Jupiter with a mass of $2.11\pm0.08 \mjup$ and a radius of $1.00\pm0.03\rjup$, with an orbital period of 1.43 days ($T_{\rm eq} = 1059 \pm 20\,\mathrm{K}$). TOI-4666 hosts a $0.70 \pm 0.06\mjup$ warm Jupiter ($T_\mathrm{eq} = 713 \pm 14\,\mathrm{K}$) with a radius of $1.11\pm0.04\rjup$, with an orbital period of 2.91 days. At a population level, we identify a decrease in planetary mass with spectral type, where late M dwarfs host less massive giant planets than early M dwarfs. More massive gas giants that deviate from this trend are preferentially hosted by more metal-rich stars. Furthermore, we find an increased binarity fraction among low-mass stars hosting gas giants, which may play a role in enhancing giant planet formation around low-mass stars.}
  {These mass characterizations contribute to the growing catalogue of well-defined giant exoplanets around low-mass stars. The observed population trends agree with theoretical predictions, where higher metallicity can compensate for lower disk masses, and wide binary systems may influence planet formation and migration through Kozai–Lidov cycles or disk instabilities.}

   \maketitle
   
\section{Introduction}
Jupiter-mass planets are scarce around M dwarfs, and theoretical models for a time predicted that stars with masses below 0.5~$M_\odot$ would have none \citep{burn_new_2021}. Both core accretion and gravitational instability - the leading mechanisms for giant planet formation - are hindered by the low disk surface densities, extended disk orbital timescales, and generally low disk masses around M dwarfs \citep{laughlin_core_2004, ida_toward_2005}. The combination of limited material and slow disk orbital motion hinders the accumulation of solids and gas, preventing cores from reaching the critical mass needed for rapid gas accretion (core accretion) or for the disk to fragment (gravitational instability), thus limiting giant planet formation before the disk disperses.

However, an increasing number of giant planets are being detected around low-mass stars \citep[e.g.][]{marcy_planetary_1998, morales_giant_2019, parviainen_TOI-519_2021, bryant_transiting_2025}. Recent studies based on photometric data from the Transiting Exoplanet Survey Satellite \citep[TESS;][]{ricker_transiting_2014}, show an occurrence rate of $0.27 \pm 0.09~\%$ for the stellar mass range $0.45-0.65~M_\odot$ \citep{gan_occurrence_2023}, and demonstrate that giant planets exist around stars with $M_\star \leq 0.4~M_\odot$ \citep{bryant_occurrence_2023}. Radial-velocity surveys consistently show that short-period ($1-10$ d), Jupiter-mass planets are uncommon but exist around M dwarfs, with recent studies placing their occurrence rate below about $2~\%$ \citep[e.g.][]{ribas_carmenes_2023, bonfils_harps_2013, pinamonti_hades_2022, pass_mid--late_2023, mignon_radial_2025}. These findings raise the question of whether a stellar mass threshold exists below which giant planet formation becomes impossible. Observational biases have prevented us from clearly identifying the point at which giant planet formation starts to decline, as faint M dwarfs ($V \geq 14$ mag) are generally not observable with optical ground-based spectrographs. With the advent of near-infrared (nIR) spectrographs like the Near-InfraRed Planet Searcher \citep[NIRPS;][]{bouchy_nirps_2025}, radial velocity (RV) surveys can now be extended to fainter M dwarfs that were previously less accessible.

However, the nIR introduces challenges not as strongly encountered in the visible. At the faint end, limitations from telluric absorption and emission lines arise, especially when the Barycentric Earth Radial Velocity (BERV) is similar to the star's systemic velocity, causing the telluric and stellar lines to overlap. Giant planets around low-mass stars induce significant RV variations due to their stronger gravitational pull, allowing observations of fainter stars than those targeted in searches for less massive planets. As a by-product, this makes them ideal targets to assess NIRPS performance on faint M dwarfs and for better understanding its faint-end limits. 

To determine the stellar mass threshold at which giant planet formation declines, an ongoing NIRPS Guaranteed Time Observations (GTO) subprogram is characterizing gas giants and brown dwarfs transiting low-mass stars to expand the known sample. 

In this paper, we confirm and characterize two gas giants transiting late-K/early-M dwarfs, contributing new mass measurements to this still relatively unexplored population. We present the ongoing subprogram (Sec.~\ref{sec:Program}), and propose a new method to improve precision for faint nIR observations affected by telluric contamination (Sec.~\ref{sec:telluric_contamination}). The first two gas giant confirmations of the program are detailed in Secs.~\ref{sec:Observations}, \ref{sec:Stellar_Properties}, and \ref{sec:Orbital_Solutions}, covering the observations, stellar characterization, and orbital solutions, respectively.

\section{NIRPS GTO \textit{TESS} Follow-Up: Gas giAnts Transiting lOw-mass Stars (GATOS)}
\label{sec:Program}
As part of the ongoing NIRPS GTO this program aims to characterize giant planets identified by the Transiting Exoplanet Survey Satellite \citep[\textit{TESS};][]{ricker_transiting_2014} orbiting M dwarfs. Observations began with the start of NIRPS operations in April 2023, with some targets having been observed during the commissioning phase to test the instrumental magnitude limits. The target selection is summarized as follows:
\begin{enumerate}
    \item We search for all \textit{TESS} objects of interest (TOIs) with a planetary radius $7.5\rearth < R_\mathrm{pl} <16 \rearth$ 
    that orbit M dwarfs, defined as stars with effective temperatures $2310~{\rm K}<T_\mathrm{eff} < 3930~{\rm K}$
    \footnote{These values were obtained from Table 5 of \citet{pecaut_intrinsic_2013}.}, resulting in 193 TOIs. The lower radius limit excludes strongly inflated Neptunes that produce small RV signals and are hard to characterize around faint stars. As the stars are cool, strongly inflated gas giants are not expected; the upper limit therefore excludes eclipsing binaries around low mass stars.
    \item We exclude 33 targets that are previously characterized planets or brown dwarfs, as well as 49 False Positive (FP) signals, and 2 False Alarms. The FPs are identified by having a \textit{TESS} Follow-up Observing Program (TFOP) Working Group (WG) disposition of Ambiguous Planet Candidates, Eclipsing Binaries (EB), or Nearby Planet Candidates (i.e., transits associated with the `wrong' star). We only exclude cases where a binary causes the transit; binary star systems hosting a transiting planet are included.
    \item Of the remaining 109 stars, 54 are observable from La Silla Observatory ($\delta \leq +20^{\circ}$), of which just 14 meet the magnitude constraints we impose for NIRPS follow-up ($J \leq 12.5$ and $H \leq 12.5$). These magnitudes roughly correspond to the limits at which $\sim$10-20  40-minutes NIRPS exposures can yield a 5$\sigma$ mass characterization of short-period gas giants.
    \item These 14 were then individually vetted through inspection of their light curve and the notes from the \textit{TESS} Spectroscopic Follow-Up WG. We exclude potential EBs, indicated by odd-even transit depth differences, secondary eclipses, or wavelength-dependent transit depth variations. V-shaped transits are kept, as they might reflect a grazing planetary transit. Such transits are more likely when larger companions transit small-sized stars. In addition to the vetting performed by \textit{TESS}, we visually vet the light curves before including targets in our program. Additionally, if another facility or team is planning high-precision spectroscopic follow-up observations, we do not initiate our own. In total 3 of the 14 are rejected, and 2 studied by other teams.
\end{enumerate}
The number of stars reported above reflects the coordinated observations table of the \textit{TESS} Spectroscopic Follow-up WG downloaded on 12 August 2025. These values have evolved since the 2023 start of our program: in January 2023 77 TOIs matched our planetary radius and effective temperature criteria, which has since increased by 116 thanks to the continuous effort of the Quick Look Pipeline (QLP) Faint search team \citep{kunimoto_tess_2022}. Some of the targets now listed as FPs were identified by our program, such as the spectroscopic binaries TOI-2341 and TOI-5295 (see App.~\ref{sec:False_Positives}). The program observed some fast-rotating stars which may still host planets, but their broadened CCFs prevent precise RV determination (see App.~\ref{app:Fast_Rotators}).

\begin{figure}
    \includegraphics[width=\linewidth]{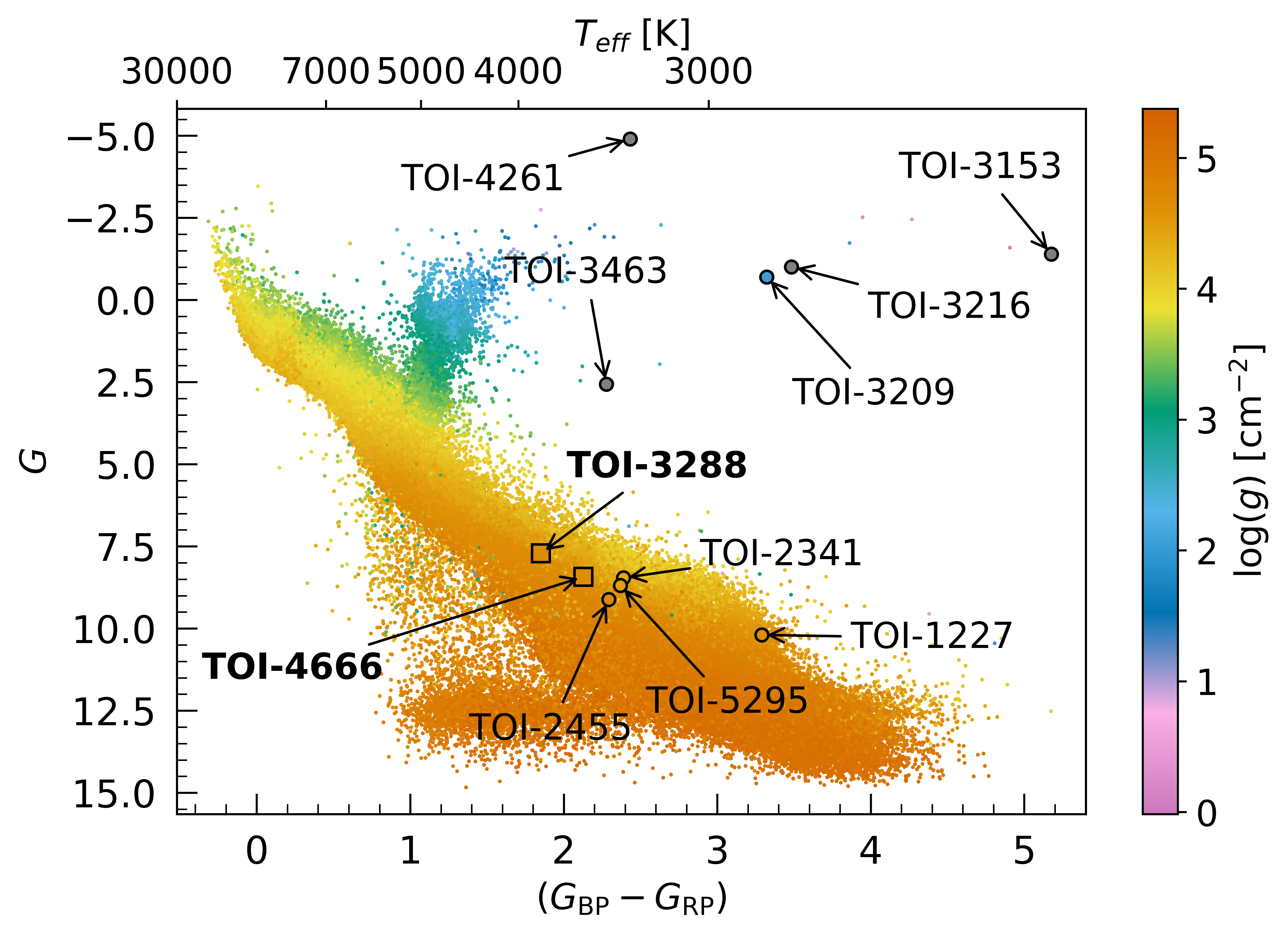}
    \caption{HR diagram of all \textit{Gaia} DR3 nearby stars with a parallax $\pi \geq 5$ mas, using the broad-band $G$ magnitude versus the colour $G_{\rm BP}$ (blue) minus $G_{\rm RP}$ (red). The colours indicate $\log(g)$, stars without a $\log (g)$ measurement are shown in grey. The six targets presented in this paper as part of the NIRPS-GTO giants subprogram are overplotted (black outlined circles), along with five stars identified as giant stars using this method. TOI-3288 and TOI-4666 (black outlined squares), hosting gas giants, are visible on the main sequence. This figure can be generated using \texttt{Gaia-HR}, available at \url{https://github.com/ygcfrensch/Gaia-HR}.}
    \label{fig:HRgaia}
\end{figure}

To vet against stars that have evolved away from the main sequence, we plot the targets of this subprogram on the \textit{Gaia} DR3 \citep{gaia_collaboration_gaia_2023} HR diagram for nearby stars ($d \leq 200\,\mathrm{pc}$; Fig.~\ref{fig:HRgaia}). The two stars discussed in this paper are on the main sequence, while some now discarded targets turned out to be evolved giant stars (see App.~\ref{app:Giant_Stars} for more details on why these are excluded from the sample).

We first obtain two spectroscopic observations for all of our targets, to promptly reject eclipsing binaries identifiable by large RV variations and/or multiple components in their cross-correlation function (CCF). We increase the observation frequency if no binarity is observed.

The NIRPS Exposure Time Calculator (ETC)\footnote{\url{https://github.com/ngrieves/NIRPS-ETC}} is used to estimate the precision achievable under optimal conditions, helping to approximate the number of observations required to achieve a $\sim 5\,\sigma$ mass characterization for planets of $\gtrsim 0.5\mjup$. The ETC is based on commissioning data taken during nominal weather conditions; therefore, it provides optimistic predictions.

This work reports on the first two targets (TOI-3288 and TOI-4666) for which the RVs confirm the presence of gas giants.

\section{Observations}
\label{sec:Observations}
\subsection{\textit{TESS} photometry}
\label{sec:TESSphotometry}
        \begin{table}
            \centering
        \caption{Properties of the \textit{TESS} extracted light curves.}
        \label{table:TESS-QLP}
            \begin{tabular}{l c c c c c} \hline\hline
            TOI & Sector & Start date & End date & $t_\mathrm{exp}$ & $\sigma_\mathrm{OOT}$\\
            & & & & [s] & [ppm]\\ \hline 
3288 & 13 & 19-06-2019 & 17-07-2019 & 1800 & 2458 \\ 
     & 27 & 05-07-2020 & 30-07-2020 & 600 & 3848 \\ 
     & 67 & 01-07-2023 & 28-07-2023 & 200 & 6349 \\ 
4666 & 31 & 22-10-2020 & 18-11-2020 & 600 & 5077 \\ \hline
        \end{tabular}\begin{flushleft}
        \textbf{Notes:} Here, $\sigma_\mathrm{OOT}$ represents the standard deviation of the out-of-transit flux in the activity-filtered light curves.\end{flushleft}
    \end{table}

TOI-3288.01 was alerted on 4 June 2021 \citep{guerrero_tess_2021} by the \textit{TESS} Science Office (TSO) after the FAINT transit search pipeline \citep{kunimoto_searching_2021} detected the planet candidate , using the Quick Look Pipeline \citep[QLP;][]{huang_photometry_2020, huang_tess_2020} on the Full Frame Image (FFI) data from sectors 13 and 27. The SPOC transit search pipeline \citep{jenkins_impact_2002, jenkins_transiting_2010, jenkins_kepler_2020} later measured the transit signature in higher cadence data from sector 67. Similarly, the FAINT pipeline detected the candidate TOI-4666.01 using the QLP FFI light curve from sector 31. After its standard series of automated and manual vetting steps \citep[e.g.][]{2018PASP..130f4502T,2019PASP..131b4506L}, the TSO issued a TOI alert on 21 December 2021. Table \ref{table:TESS-QLP} presents the sectors in which \textit{TESS} observed TOI-3288 and TOI-4666 together with the photometric cadences and out-of-transit scatters. \textit{TESS}'s simple aperture photometry (SAP) is affected by instrumental systematic effects mostly originating from the spacecraft pointing, which have been shown to be effectively corrected through the PDC algorithm originally developed for the \textit{Kepler} mission \citep[see][]{stumpe_kepler_2012, stumpe_multiscale_2014, smith_kepler_2012, kinemuchi_demystifying_2012}. The PDC-corrected SAP (i.e. PDCSAP) is delivered by the SPOC and \textit{TESS}-SPOC pipelines for a series of \textit{TESS} sectors. Unfortunately, for our targets, only sector 67 for TOI-3288 has available PDCSAP data. Therefore, we consistently extracted our own PDCSAP photometry from the FFIs \citep[which we accessed through the \texttt{lightkurve} Python package;][]{lightkurve_collaboration_lightkurve_2018} in a similar manner to the \textit{TESS}-SPOC pipeline. First, we excluded any FFI flagged for quality issues, recognizable by a non-zero quality parameter. Then, we built the photometric apertures. As both stars are in crowded fields, we used the \texttt{TESS-cont} algorithm\footnote{Available at \url{https://github.com/castro-gzlz/TESS-cont}} \citep{castro-gonzalez_toi5005_2024} to create apertures that minimize the flux from nearby contaminant stars, restricting them to those pixels where at least $60~\%$ of the flux originates from the target star (see App.~\ref{app:TESS-cont} for the selected apertures). We subsequently obtained the SAP photometry by subtracting the background flux, which we obtained by defining a background aperture as the nearby pixels with median flux lower than the minimum flux plus $2.5~\%$ of the standard deviation, excluding all pixels which belong to the aperture mask. Finally, we applied the Cotrending Basis Vectors (CBV) corrector implemented in \texttt{lightkurve} and corrected from crowding through the CROWDSAP and FLFRCSAP metrics computed by \texttt{TESS-cont} for the selected apertures. (the photometric data are presented in App.~\ref{app:photometry}).

\subsection{Zorro Speckle Interferometry}
If a star hosting a planet candidate has a close bound binary companion or  a background star that is angularly close in the line of sight, the additional light falling within the photometric aperture can create a FP exoplanet detection due to the effect on the transit depth \citep[e.g.][]{ciardi_understanding_2015,2017A&A...606A..75C} or corrupt the determined parameters for both the planet and host star \citep[e.g.][]{furlan_densities_2017, furlan_unresolved_2020,2022MNRAS.509.1075C}. We could obtain high-resolution optical speckle imaging observations to search for close-in companions unresolved in \textit{TESS} or other follow-up observations for TOI-3288, but not for TOI-4666. 

\begin{figure}
    \includegraphics[width=\linewidth]{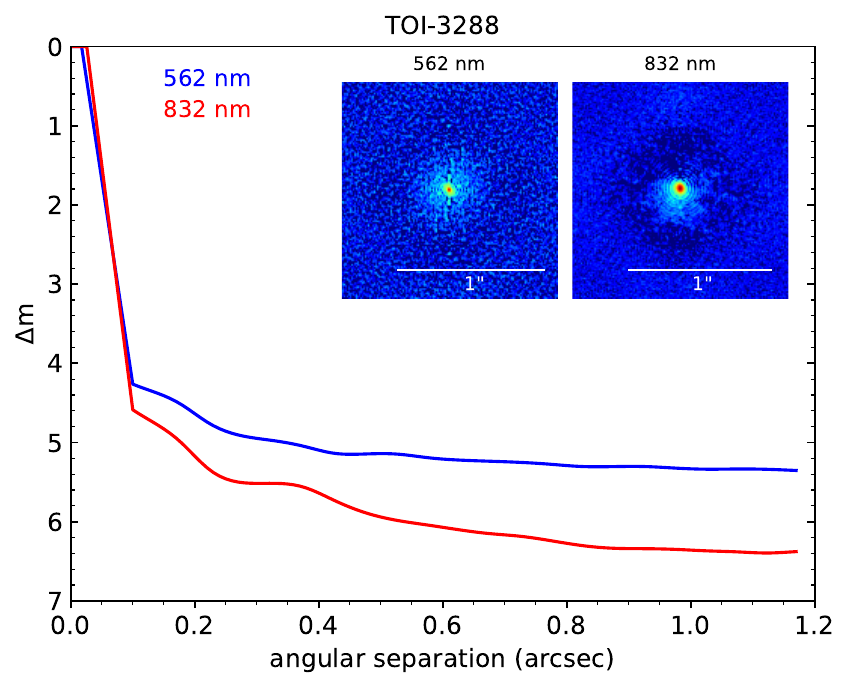}
    \caption{The $5\,\sigma$ magnitude contrast curves as observed by Zorro in both filters as a function of the angular separation out to $1.2\arcsec$. The insets show the reconstructed speckle images of TOI-3288 with a $1\,\arcsec$ scale bar. TOI-3288 was found to have no close companions from the diffraction limit ($0.02\,\arcsec$) out to $1.2\,\arcsec$ to within the magnitude contrast levels achieved.}
    \label{fig:zorro_toi3288}
\end{figure}

TOI-3288 was observed on 2022 May 17 UT using the speckle instrument Zorro on the Gemini South 8-m telescope \citep{scott_twin_2021}. Zorro provides simultaneous speckle imaging in two bands (562 nm and 832 nm) with output data products including a reconstructed image with robust contrast limits on companion detections. Thirteen sets of $1000 \times 0.06$ sec exposures were collected and subjected to Fourier analysis in our standard reduction pipeline \citep{howell_speckle_2011}. Figure~\ref{fig:zorro_toi3288} shows our final $5\,\sigma$ magnitude contrast curves and the reconstructed speckle images. We find that TOI-3288 does not have a stellar companion brighter than 4.5 to 6.4 magnitudes below that of the target star from the diffraction limit (20 mas) out to $1.2\,\arcsec$. At the distance of TOI-3288 ($d=200\,\mathrm{pc}$) these angular limits correspond to spatial limits of 4 to 240 AU.

Beyond the $1.2\,\arcsec$ field probed by Zorro, \textit{Gaia} reveals a stellar companion at $\sim2.3\,\arcsec$ (\textit{Gaia} DR3 6685431748040148992) with $G=17.2$ mag, which is discussed in more detail in Sec.~\ref{sec:Binarity}.

\subsection{Ground-based photometry}
\label{subsec:ground}
To determine the true source of the \textit{TESS} detection, we acquired ground-based time-series follow-up photometry of the fields around TOI-3288 and TOI-4666 as part of the \textit{TESS} Follow-up Observing Program \citep[TFOP;][]{collins_tess_2019}\footnote{\url{https://tess.mit.edu/followup}}. We used the {\tt \textit{TESS} Transit Finder}, which is a customized version of the {\tt Tapir} software package \citep{jensen_tapir_2013}, to schedule our transit observations. The light curve data are available on the {\tt EXOFOP-TESS} website\footnote{\url{https://exofop.ipac.caltech.edu/tess}\label{foot:EXOFOP}} and are included in the global modelling described in Sec.~\ref{sec:Orbital_Solutions}. TOI-3288 was observed with LCOGT and TOI-4666 with ExTrA, as part of the coordinated photometric follow-up to validate the transit, depending on the availability of the facilities.

\subsubsection{LCOGT -- TOI-3288}
\label{sec:LCOGT}
The Las Cumbres Observatory Global Telescope \citep[LCOGT;][]{brown_cumbres_2013} 1.0\,m and 0.4\,m network nodes used in the TOI-3288 observations are located at Cerro Tololo Inter-American Observatory in Chile (CTIO), Siding Spring Observatory near Coonabarabran, Australia (SSO), and South African Astronomical Observatory near Cape Town South Africa (SAAO). The 1\,m telescopes are equipped with $4096\times4096$ SINISTRO cameras which have an image scale of $0.389\arcsec$ per pixel, resulting in a $26\arcmin\times26\arcmin$ field of view. The 0.4\,m telescopes are equipped with a $2048\times3072$ pixel SBIG STX6303 camera having an image scale of $0.57\arcsec$ pixel$^{-1}$, resulting in a $19\arcmin\times29\arcmin$ field of view. The standard LCOGT {\tt BANZAI} pipeline \citep{mccully_real-time_2018} calibrated all LCOGT images, and differential photometric data were extracted using {\tt AstroImageJ} \citep[{\tt AIJ};][]{collins_astroimagej_2017}. All observations of TOI-3288 are affected by contamination from the stellar companion \textit{Gaia} DR3 6685431748040148992.

\subsubsection{ExTrA -- TOI-4666}
\label{sec:ExTrA}
ExTrA \citep[Exoplanets in Transits and their Atmospheres,][]{bonfils_extra_2015} is a low-resolution near-infrared (0.85 to 1.55~$\mu$m) multi-object spectrograph fed by three 60-cm telescopes located at La Silla Observatory in Chile. A total of 10 transits of TOI-4666 b were observed using two (2 transits) or three (8 transits) of the ExTrA telescopes. We used $8\arcsec$ diameter aperture fibres and the low-resolution mode ($R\sim20$) of the spectrograph, with an exposure time of $60$ seconds. Five fibres are positioned in the focal plane of each telescope to select light from the target and four comparison stars. The resulting ExTrA data were analysed using custom data reduction software \citep{cointepas_toi-269_2021}.

\subsubsection{ASAS-SN}
\label{sec:asassn}
We complemented the dataset using observations from the All Sky Automated Survey for SuperNovae (ASAS-SN; \citealt{Kochanek2017}), an automated photometric programme that looks out for supernovae and other transient events. ASAS-SN is a network of 24 ground-based 14-centimetre telescopes distributed over six sites. Due to their cadence, these data are not used to detect transit events but instead to follow the stars' behaviour. We obtained $\sim$8 years of $g$-band photometry and $\sim$ 4.5 years of $V$-band photometry of both targets. Before analysing it, the data were nightly binned and linearly detrended from long-term variations. 

\subsection{HARPS and NIRPS spectroscopy}
        \begin{table}
            \centering
            \caption{Characteristics of the NIRPS and HARPS observations.}
            \label{table:NIRPS_HARPS}
            \begin{tabular}{l l c c c c} \hline\hline
            TOI & $N_\mathrm{meas}$ & $t_\mathrm{exp}$ & $\mathrm{med}(\sigma_\mathrm{RV})^{(a)}$ \\
            & [d] & [s] & [m s$^{-1}$] \\ \hline
TOI-3288 \\
 ~~\footnotesize{HARPS-EGGS} & 6 & 2400 & 66 \\
 ~~\footnotesize{HARPS-HAM} & 3 & 2400 & 42 \\
 ~~\footnotesize{NIRPS-HE} & 13 & $3 \times 800$ & 46 \\ \hline 
TOI-4666 \\
 ~~\footnotesize{HARPS-EGGS} & 10 & 2400 & 87 \\
 ~~\footnotesize{HARPS-HAM} & 2 & 2400 & 58 \\
 ~~\footnotesize{NIRPS-HE} & 13 & $3\times 800^{(b)}$ & 40 \\ \hline
            \end{tabular}
            \begin{flushleft}
        \textbf{Notes:} $^{(a)}$ In the case of NIRPS, the data are binned by night.\\
        $^{(b)}$ In some cases, the system failed to obtain all three consecutive exposures; the successful sub-exposures are included in the analysis (see Table~\ref{tab:TOI-4666_RVs} for details).
        \end{flushleft}
        \end{table}
NIRPS is an Adaptive Optics (AO)-assisted and fibre-fed high-resolution spectrograph operating at the ESO 3.6-meter telescope in La Silla Observatory. It had its first light in May 2022 \citep[][]{wildi_first_2022, artigau_nirps_2024, bouchy_nirps_2025}. NIRPS observes the nIR ($\lambda=0.98-1.8\,\upmu$m). Fibre modal noise is more dominant and can severely limit the highest achievable signal-to-noise ratio on bright stars \citep[S/N;][]{iuzzolino_preliminary_2014}. NIRPS combines fibre stretchers, octagonal fibres, double scramblers, and AO tip-tilt scrambling \citep{blind_modal_2022, frensch_nirps_2022} to significantly reduce modal noise. Simultaneous observations are possible with HARPS \citep[High-Accuracy Radial Velocity Planetary Searcher;][]{mayor_setting_2003}, which covers the optical range ($\lambda = 0.38-0.69\,\upmu$m).

NIRPS uses a high-order AO system to couple starlight into four relatively small multi-mode fibres (29-$\upmu$m and 66-$\upmu$m). There are two observing modes: High-Accuracy (HA) and High-Efficiency (HE), both of which include one science fibre and one reference fibre. For the faint targets in the giants program, we use the HARPS-EGGS (HE) and NIRPS-HE modes to maximize throughput and minimize modal noise. During the period that the EGGS shutter was malfunctioning ($\sim$June to late September 2024), we performed observations in HARPS-HAM (HA) mode instead. The reference fibre for NIRPS is pointed at the sky; these simultaneous sky observations are used to correct for telluric lines. As both instruments are ultra-stable, continuous drift monitoring is not required.

The stellar spectra for HARPS and NIRPS are extracted using version 3.2.0 of the NIRPS Data Reduction Software (DRS) adapted from the ESPRESSO DRS \citep{pepe_espresso_2021}. For NIRPS, the DRS includes corrections for telluric absorption and emission \citep[][Srivastava et al., A\&A, in review]{allart_automatic_2022}. To correct telluric absorption, a synthetic spectrum is created. First, the CCF is computed to obtain an average telluric line per molecule. Using physical parameters from the HITRAN database \citep{gordon_hitran2020_2022}, a model is fitted to each average line. The fitted parameters are then used to generate the synthetic telluric spectrum, which is subtracted from the observations. Telluric emission is corrected by scaling lines from the reference fibre to the science fibre using ratios determined from previous sky–sky observations. For HARPS, the DRS does not correct for telluric contamination as the optical domain is less affected. These corrections are sufficient to enable sub-m/s precision for bright targets, such as Proxima \citep{suarez_nirps_2025}; however, fainter targets require some additional cleaning to improve precision and accuracy. In Sec.~\ref{sec:telluric_contamination} we discuss the technique used to further reduce telluric contamination.

The line-by-line \citep[LBL; ][]{dumusque_LBL_2018, artigau_line-by-line_2022} method for determining RV measurements has been optimised to handle spectral outliers and to minimise their impact. Radial velocities are measured independently in all valid lines (typically $\sim$16\,000 in NIRPS), and an error-weighted average is computed that accounts for outliers through a finite-mixture model. This produces a soft clipping (a gradually decreasing weight) at the $4$–$5\,\sigma$ level. In the presence of \textit{uncorrelated} outliers, this method significantly outperforms CCF measurements \citep[see Figures~5 and 6 in][]{artigau_line-by-line_2022}.

A key limiting case arises when numerous lines are distorted in a correlated manner, at low-enough significance to avoid rejection by the finite-mixture model, yet in sufficient numbers to significantly bias the combined RV. This is particularly the case in low-SNR observations of targets that have a low systemic velocity, which is the case here. Since all water lines are affected in a correlated way by finite-resolution effects \citep[see Figures~4 and 5 in][]{wang_characterizing_2022}, there is a bulk effect that cannot be mitigated by rejecting individual outliers. This occurs when water absorption lines overlap within a few km/s with water lines in cool dwarfs, i.e., when the sum of the barycentric correction and the systemic velocity of the star is within about one resolution elements of zero (i.e., $<$5\,km/s). Not all observed stars are affected, as this depends on their systemic velocity and ecliptic latitude. 

This correlated effect must be addressed either through improved telluric correction (i.e., better accounting of finite-resolution effects and refined telluric line profiles) or through parametrisation of its impact on the mean line profile \citep[see][for a strategy where OH lines are removed]{parc_TOI-756_2025}. While the former approach is the focus of ongoing work, in the present analysis we opted for the latter. 

We concentrate on post-processing the CCF RVs to develop CCF-specific mitigation techniques, complementing the available LBL strategies. Table~\ref{table:NIRPS_HARPS} summarises the characteristics of the post-processed observations, and the individual RVs are presented in App.~\ref{app:RVs}. The NIRPS RVs were obtained in three consecutive 800 s sub-exposures to reduce instrumental systematics. Since the individual sub-exposures do not reach the required precision, they are combined into a single RV measurement, which we refer to as ‘night-binning’ throughout the paper.

\section{Stellar properties}
\label{sec:Stellar_Properties}
Table~\ref{table:stellar_parameters} summarizes the stellar parameters of TOI-3288 and TOI-4666. The TIC working group \citep{paegert_tess_2021, stassun_revised_2019} magnitudes are reported. The $J$, $H$, and $K$ magnitudes follow from 2MASS \citep{cutri_2MASS_2003}. The $V$ magnitudes are calculated from the \textit{Gaia} colours ($G$, $B'$, $R'$) and the $B$ magnitudes are calculated following \cite{stassun_TIC_2018}, as the stars are too faint to be included in the \textit{Tycho} catalogue \citep{esa_hipparcos_1997}. 
Surface gravities $\log g$, parallaxes $\pi$, and corresponding distances $d$ originate from \textit{Gaia} DR3 \citep{gaia_collaboration_gaia_2023}. Effective temperatures $T_\mathrm{eff}$ and abundances $[\mathrm{X/H}]$ are determined through a spectral analysis of NIRPS and HARPS data to ensure consistency between the independently obtained values; see Sec.~\ref{sec:Spectral_Analysis} for details. Spectral types are inferred from the NIRPS $T_\mathrm{eff}$ using Table 5 of \cite{pecaut_intrinsic_2013}. The extinction $A_\mathrm{V}$ and bolometric luminosity $L_\mathrm{bol}$, are derived from an SED fit (Sec.~\ref{sec:SED_Analysis}). From empirically appropriate relations, the stellar radius and mass are estimated as described by \cite{mann_how_2015, mann_how_2019} for $T_\mathrm{eff} < 4000\,\rm{K}$. For TOI-3288 the $K$ band is contaminated by the nearby stellar companion which is 16 times fainter than the primary, the contamination is low, but impacts the inferred $R_{\star}$ and $M_{\star}$.
The \textit{TESS} light curves show possible rotational modulations, which allow us to estimate rotation period $P_\mathrm{rot}$ (Sec.~\ref{sec:Prot_TESS}).

\begin{table*}
\caption{Stellar parameters of the stars presented in this paper.}
\label{table:stellar_parameters}

\centering
\begin{tabular}{l l  c c c c}
\hline\hline
&  & TOI-3288 & TOI-4666 & &\\ 
\multicolumn{2}{l}{Parameter}  & \footnotesize{TIC 79920467} & \footnotesize{TIC 165202476} & Source & Sec. \\ \hline
SpType &  & K9V & M2.5V & $T_\mathrm{eff}$ NIRPS & \ref{sec:Stellar_Properties}\\
$B$ & [mag]  & $16.5\pm 0.1$  & $16.75\pm 0.08$  & \textit{TESS}\\
$V$ & [mag]  & $15.0\pm 0.2$  & $15.3\pm 0.2$  & \textit{TESS}\\
$G$ & [mag]  & $14.2$ & $14.4$ & \textit{Gaia} DR3\\
$J$ & [mag]  & $12.04\pm 0.03$  & $11.98\pm 0.02$  & 2MASS\\
$H$ & [mag]  & $11.42\pm 0.03$  & $11.29\pm 0.02$  & 2MASS \\
$K$ & [mag]  & $11.20\pm 0.03$  & $11.09\pm 0.03$  & 2MASS\\
$\pi$ & [mas]  & $4.97\pm 0.02$  & $6.49\pm 0.02$  & \textit{Gaia} DR3\\
$d$ & [pc]  & $201.3\pm 1.0$  & $154.1\pm 0.4$  & $\pi$ & \ref{sec:Stellar_Properties}\\
$\log(g)$ & [cm s$^{-2}$]  & $4.658_{-0.004}^{+0.003}$  & $4.626_{-0.004}^{+0.008}$  & \textit{Gaia} DR3\\
$T_{\mathrm{eff}}$ & [K]  & $3933\pm 48$  & $3512\pm 36$  & NIRPS & \ref{sec:NIRPS_spec}\\
$T_\mathrm{eff}$ & [K] & $3881 \pm 111$ & $3720 \pm 103$ & HARPS & \ref{sec:HARPS_spec}\\
$\mathrm{[M/H]}^{(a)}$ & [dex]  & $-0.1\pm 0.1$  & $-0.1\pm 0.1$  & NIRPS & \ref{sec:NIRPS_spec}\\
$\mathrm{[Fe\,/H]}$ & [dex] & $0.27\pm 0.13$ & $0.19\pm 0.12$ & HARPS & \ref{sec:HARPS_spec}\\
$\mathrm{[Mg\,I/H]}$ & [dex]  & $0.10\pm 0.09$  &  & NIRPS & \ref{sec:NIRPS_spec}\\
$\mathrm{[Si\,I/H]}$ & [dex]  & $0.16\pm 0.14$  &  & NIRPS & \ref{sec:NIRPS_spec}\\
$\mathrm{[Ca\,I/H]}$ & [dex]  & $0.03\pm 0.21$  & $0.3\pm 0.5$  & NIRPS & \ref{sec:NIRPS_spec}\\
$A_V$ & [mag]  & $0.10\pm 0.09$  & $0.02\pm 0.02$  & SED & \ref{sec:SED_Analysis}\\
$L_\mathrm{bol}$ & [$L_{\odot}$]  & $0.110\pm 0.003$  & $0.060\pm 0.003$  & SED & \ref{sec:SED_Analysis}\\
$R_{\star}$ & [$R_{\odot}$]  & $0.67 \pm 0.02$ & $0.59 \pm0.02$ & \cite{mann_how_2015} & \\
$M_{\star}$ & [$M_{\odot}$]  & $0.64\pm 0.01$  & $0.58 \pm 0.01$ & \cite{mann_how_2019} & \\
$P_{\mathrm{rot}}$ & [d]  & $\sim 13.7^{(b)}$ & $\sim 15.5$  & \textit{TESS} & \ref{sec:Prot_TESS}\\
$P_{\rm rot}$ & [d] & $16.3$ & $15.3$ & ASAS & \ref{sec:Prot_TESS} \\
\hline
\end{tabular}
\begin{flushleft}\footnotesize
\textbf{Notes:} $^{(a)}$ The [M/H] values reported here are derived from spectral metallicity and should be considered lower bounds for the star's overall metallicity. They reflect primarily the OH abundance and should not be directly compared to integrated [M/H] measurements from other sources.\\
$^{(b)}$ Ambiguous variation.
\end{flushleft}
\end{table*}

\subsection{Spectral analysis}
\label{sec:Spectral_Analysis}
\subsubsection{NIRPS}
\label{sec:NIRPS_spec}
For each star, the spectroscopic stellar parameters are derived from a single high-resolution template spectrum, obtained by combining the individual telluric-corrected spectra from the APERO DRS \citep{cook_apero_2022}. We note that the resulting templates have low S/N (54 and 67 for TOI-3288 and TOI-4666, respectively). Following the methodology of \citet{jahandar_comprehensive_2024, jahandar_chemical_2025}, we first determine the effective temperature $T_\mathrm{eff}$ and overall metallicity [M/H] by fitting groups of spectral features to a pregenerated grid of PHOENIX ACES stellar models \citep{husser_new_2013}. The models are convolved to the resolution of NIRPS and have $\log{g}=4.50$, as both stars have surface gravities within 0.2\,dex of this value (see Table \ref{table:stellar_parameters}) where variations in $\log g$ do not affect the spectral analysis. We find $T_\mathrm{eff}=3933\pm48$\,K and $\mathrm{[M/H]}=-0.1\pm0.1$\,dex for TOI-3288 while the TOI-4666 spectrum yields $T_\mathrm{eff}=3512\pm36$\,K and $\mathrm{[M/H]}=-0.1\pm0.1$\,dex. We note that the effective temperature of TOI-3288 classifies it as a K9. This discrepancy arises because the \textit{TESS} Input Catalog \citep[TICv8;][]{stassun_revised_2019} often relies on empirical color-based relations, whereas the spectral analysis presented here is based on direct observations.

To measure the abundances of individual chemical species, we then perform fits on individual spectral lines, using a PHOENIX grid with fixed $T_\mathrm{eff}$ of 3940\,K and 3520\,K for TOI-3288 and TOI-4666, respectively \citep{jahandar_comprehensive_2024, jahandar_chemical_2025}. Due to the low S/N of the templates, we were only able to fit very few lines per element; all reported abundances are based on at least three lines (see Table~\ref{table:stellar_parameters}). To determine the uncertainty on the abundance measurements, we take the dispersion of the individual measurements. We note that the overall metallicities reported for both stars ($\mathrm{[M/H]}=-0.1\pm0.1$\,dex; see Table~\ref{table:stellar_parameters}) are values typically considered as a first approximation of [M/H]. Being measured directly from the spectrum, this [M/H] is affected by the relative number of lines of the different chemical species. For M dwarfs in the nIR, the spectrum is often dominated by OH lines, resulting in a spectral [M/H] that can differ from one measured at optical wavelengths, where Fe lines are more prominent. To account for this difference, we typically would report an \textit{integrated} [M/H], defined as an average of the abundance of each element \citep{jahandar_comprehensive_2024}. Because of the low S/N and small number of measured chemical species (3 for TOI-3288 and only 1 for TOI-4666, see Table~\ref{table:stellar_parameters}), we decided to instead report the [M/H] based on the overall stellar spectrum.

\subsubsection{HARPS}
\label{sec:HARPS_spec}

In addition, we derived $T_\mathrm{eff}$ and metallicity $[\mathrm{Fe/H}]$ (i.e. iron abundance) from the HARPS spectrum using the machine learning tool {\tt ODUSSEAS}\footnote{\url{https://github.com/AlexandrosAntoniadis/ODUSSEAS}} \citep{Antoniadis20,Antoniadis24}. We first combined all the individual HARPS spectra with the task {\tt scombine} within IRAF\footnote{IRAF is distributed by National Optical Astronomy Observatories, operated by the Association of Universities for Research in Astronomy, Inc., under contract with the National Science Foundation, USA.} to obtain a higher S/N spectrum. This spectrum is used by {\tt ODUSSEAS} to measure the pseudo-EWs of more than 4000 lines to apply a machine learning model trained with the same lines, measured and calibrated on a reference sample of 47 M dwarfs observed with HARPS for which their $[\mathrm{Fe/H}]$ were obtained from photometric calibrations \citep{Neves12} and their $T_\mathrm{eff}$ from interferometric calibrations \citep{Khata21}. We note that given the low S/N of the individual spectra for TOI-4666 we only used the cleaned spectra missing the bluest and reddest orders, which causes a reduction of 13$\%$ of the number of lines measured. With this method, we derived $T_\mathrm{eff}$\,=\,3881\,$\pm$\,111\,K and $[\mathrm{Fe/H}]$\,=\,0.27\,$\pm$\,0.13\,dex for TOI-3288 and $T_\mathrm{eff}$\,=\,3720\,$\pm$\,103\,K and $[\mathrm{Fe/H}]$\,=\,0.19\,$\pm$\,0.12\,dex for TOI-4666. The combined measurements from HARPS and NIRPS suggest that both stars are metal-rich but it is hard to have a good constraint on the real metallicity given the low S/N of the spectra. This prevents the derivation of an accurate temperature for TOI-4666, as shown by the $\sim 2~\sigma$ discrepancy between the value found with respect to the one obtained from the NIRPS spectrum. Therefore, these parameter values have to be taken with care.

\subsection{SED analysis}
\label{sec:SED_Analysis}
The bolometric flux $F_\mathrm{bol}$ and extinction $A_\mathrm{V}$ as reported in Table~\ref{table:stellar_parameters} are obtained from an independent broadband spectral energy distribution (SED) analysis with the \textit{Gaia} DR3 parallaxes \citep[no systematic offset is applied; see, e.g.,][]{stassun_parallax_2021}. The $JHK_S$ magnitudes follow from 2MASS, the W1--W3 magnitudes from WISE, the $G_{\rm BP}$, $G_{\rm RP}$ magnitudes from \textit{Gaia}, and the absolute flux-calibrated \textit{Gaia} spectrophotometry. The available photometry spans the full stellar SED over the wavelength range of 0.4-10 $\upmu$m (see Figure~\ref{fig:sed}). 

We fitted PHOENIX stellar atmosphere models \citep{husser_new_2013} with the effective temperature $T_\mathrm{eff}$ from NIRPS, and the surface gravity $\log g$, and metallicity [Fe/H] set to the spectroscopically determined values from Sec.~\ref{sec:Spectral_Analysis}. The fit's $A_\mathrm{V}$ is limited by the maximum line-of-sight value from the Galactic dust maps of \citet{schlegel_maps_1998}. The $F_\mathrm{bol}$ at Earth is determined by integrating the unreddened model SED. Combining the $F_\mathrm{bol}$ and the parallax provides the bolometric luminosity $L_\mathrm{bol}$.

\subsection{Rotational period}
\label{sec:Prot_TESS}
The raw \textit{TESS} light curves (SAP flux) and detrended flux from the QLP light curves were analysed using three different periodicity analysis techniques: (i) Lomb–Scargle periodograms \citep[e.g.,][]{scargle_studies_1982, horne_prescription_1986, press_fast_1989}, (ii) Fast Fourier Transform \citep[FFT; see][for details]{zhan_complex_2019}, and (iii) wavelet analysis \citep{grossmann_decomposition_1984}. We searched for variability signatures by considering periods from the Nyquist frequency (minimum period) up to 30 days, given that each TESS sector observes the sky for $\sim27\,\mathrm{days}$. Diagnostic plots were constructed for each star, combining the light curves and results from all three methods to provide a more robust visual analysis. This enables the more accurate identification of consistent periodicities and facilitates the classification of stellar variability types. Our approach follows the same procedure described in \citet{canto_martins_search_2020}. 

As the rotation periods are similar across the various sectors and three techniques, we report weighted mean values of $\sim13.7$ days for TOI-3288 and $\sim15.3$ days for TOI-4666 (see Table \ref{table:stellar_parameters}). The value for TOI-3288 requires caution, its three sectors exhibit complex patterns, which makes it difficult to attribute the variability to a consistent astrophysical origin. Since the behaviour of the light curves is not consistent with any known modulation reported in the literature, we cannot confidently conclude that the observed variability is astrophysical. In such cases, we follow the classification scheme from \cite{canto_martins_search_2020}, where the variability is labelled as `ambiguous variation'. 

To confirm the rotation measurements obtained from the \textit{TESS} observations, we analysed the ASAS-SN data. We used a Generalised Lomb-Scargle periodogram \citep{Zechmeister2009} to evaluate the presence of periodic signals in the data. For TOI-3288 we detected a significant signal in the $g$-band data with a period of 16.3 days. The $V$-band data showed a signal at the same period, although not significant. The measurement differs from the \textit{TESS} period. However, complexity of the sector-by-sector variation of the \textit{TESS} data, combined with the fact that the sectors are spaced a year or more apart, might be obscuring the true astrophysical periodicity. For TOI-4666 we identified a non-significant signal in the $g$-band data with a period of 15.4 days, consistent with the period measured in the \textit{TESS} light curve. Figures~\ref{fig:asassn1} and \ref{fig:asassn2} shows the $g$-band data, and their respective periodograms, of TOI-3288 and TOI-4666.

\section{Telluric Mitigation and Order Selection}
\label{sec:telluric_contamination}
If the systemic velocity $v_\mathrm{sys}$ and BERV overlap, Earth's atmospheric lines blend with the stellar lines, which translates into a blended CCF containing contributions from both the RV signal and residual telluric features. The DRS telluric corrections already significantly reduce the contamination; however, fainter targets can still be heavily influenced. \cite{allart_automatic_2022} use the CCF method to compute average telluric lines, from which a reconstructed telluric spectrum is derived. At low S/N, the average telluric lines are poorly defined, which results in an inaccurate reconstructed spectrum and thus in residuals or over-corrections in the final data. The S/N reaches only $\sim 12$ in the reddest orders for both TOI-3288 and TOI-4666.

\begin{figure}
    \centering
    \includegraphics[width=\linewidth]{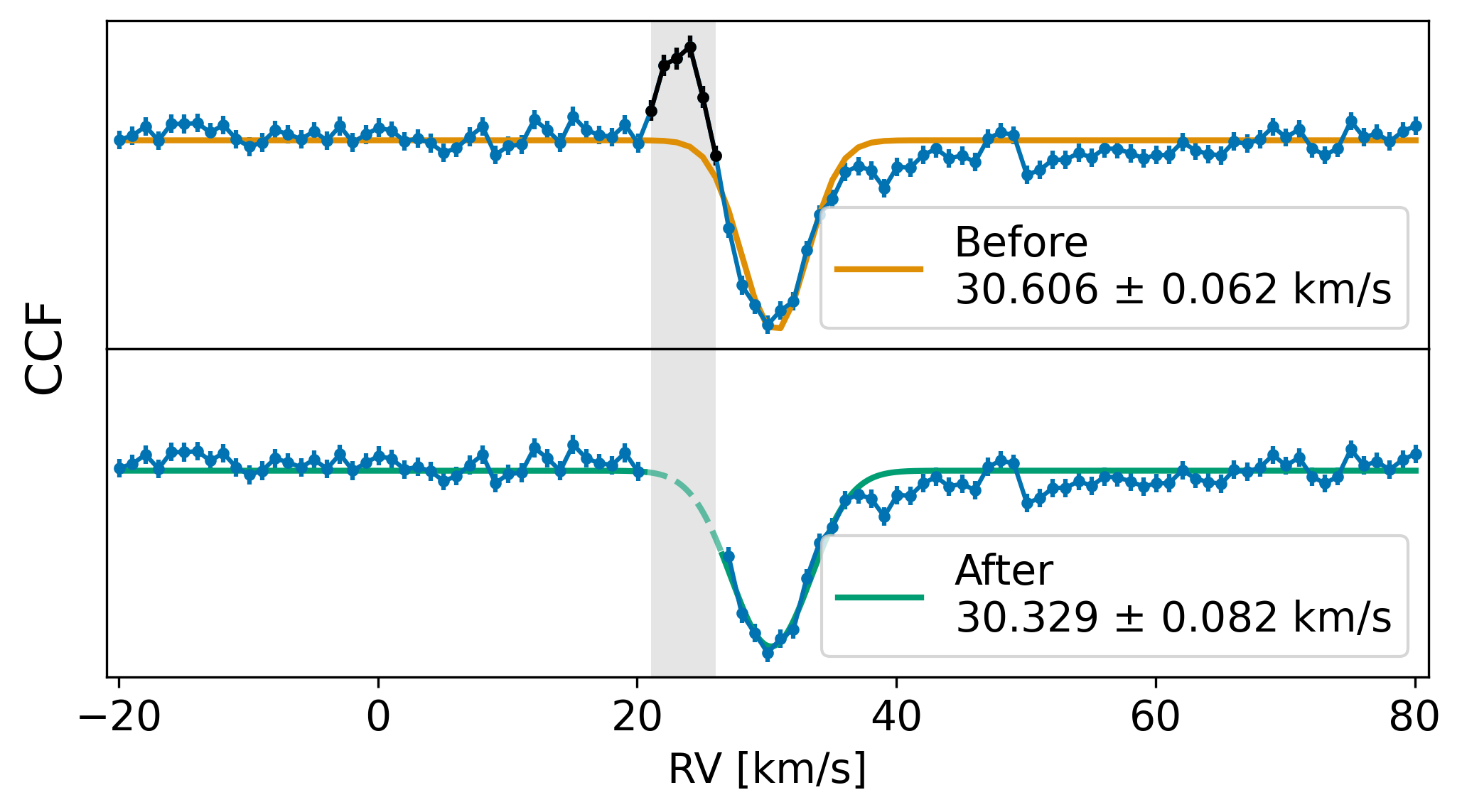}
    \caption{Normalized CCF of TOI-3288. The upper panel shows the full CCF, while the lower panel shows the CCF with the BERV window ($\pm \frac{1}{2}{\rm FWHM}$) excluded. The Gaussian fit is indicated in orange (top) and green (bottom). The uncorrected RV is red-shifted by $\sim300$ m/s.}
    \label{fig:BERV_exclusion}
\end{figure}

To reduce the remnant effect, we introduce an additional post-processing step. First, the CCFs are computed similarly to the DRS using the \verb|iCCF| Python package\footnote{\url{https://github.com/j-faria/iCCF}} \citep{faria_iCCF_2025}. The spectra are cross-correlated with numerical stellar masks, closely matched to the spectral types of each TOI (M1 for NIRPS and M2 for HARPS) \citep{fellgett_concerning_1953, baranne_elodie_1996}. From the full CCF, we estimate the FWHM. Then, for each NIRPS frame, we identify the BERV and remove all CCF values within the $\rm{BERV} \pm \frac{1}{2} \rm{FWHM}$ window from the RV array. A Gaussian is fit to the remaining, BERV-excluded CCF, resulting in an uncontaminated RV value, see Fig.~\ref{fig:BERV_exclusion} for an illustration of the approach. This method works particularly well for very faint, heavily contaminated targets. However, it is not recommended for brighter stars, as significant information is lost in the process (e.g., flux from uncontaminated orders is also excluded). For the joint fit presented in Sec.~\ref{sec:Orbital_Solutions}, we use the RVs obtained from this technique.

For HARPS, another approach is used: orders 1 to 30 are excluded due to critically low S/N in the blue, and the two reddest orders are removed to avoid potential telluric contamination. The remaining orders are summed to compute the final CCF, following the standard DRS procedure.

\section{Orbital solution}
\label{sec:Orbital_Solutions}
We combine the post-processed RVs with the photometric data using the \texttt{Juliet} Python package \citep{espinoza_juliet_2019}, applying \texttt{dynesty} \citep{speagle_dynesty_2020} for dynamic nested sampling to estimate Bayesian posteriors and evidences. A joint fit is performed, combining the NIRPS and HARPS RVs with all available photometric data. The resulting orbital solution is presented in this section. We do not include a Gaussian Process (GP) regression, as the stellar rotation period is poorly constrained. In addition, we fit the HARPS and NIRPS data separately using photometric priors on the epoch and orbital period, and compare the derived semi-amplitudes to verify consistency between the optical and nIR. A mismatch could indicate a diluted eclipsing binary.

The priors used for the joint fit are presented in Table \ref{table:priors}. The prior values for the orbital period $P$, time of transit $T_0$, and stellar density $\rho_\star$, were retrieved from the ExoFOP website\footref{foot:EXOFOP}. We adopt priors with broad bounds to ensure they are relatively uninformative. The eccentricity $e$ and argument of periastron $\omega$ are either fixed ($e = 0$, $\omega = 90^\circ$), or $e$ is modelled using a beta prior following \citet{kipping_parametrizing_2013}, while $\omega$ is assigned a uniform prior between $0^{\circ}$ and $180^{\circ}$. The models are compared using the log-evidence to determine whether fixing $e$ and $\omega$, or allowing them to vary, provides a better fit to the data. A uniform prior between 0 and 1 is applied to the radius ratio ($R_\mathrm{pl}/R_\star$) and impact parameter ($b$). Since the transits are not V-shaped and thus not grazing, values of $b$ greater than 1 are outside the considered range. All other priors follow the recommendations from the \texttt{Juliet} documentation. The resulting posterior distributions are presented as \texttt{corner} plots in App.~\ref{app:Corner} \citep{foreman-mackey_cornerpy_2016}.

The integrated AO guiding camera images are available with each observation. These can be inspected for contaminating sources in the RV observations (App.~\ref{app:AO_Guiding}). The \textit{TESS} light curves are corrected for contamination. However, as TOI-3288 has a stellar companion, a dilution factor $D$ is included as a uniform prior from 0.1 to 1 for the ground-based photometry. As the ExTrA light curves of TOI-4666 are extracted with an $8\arcsec$ diameter aperture, and the closest star lies at $38\arcsec$, contamination is negligible. Therefore, the dilution factor is fixed to 1.
Each ExTrA transit and telescope is treated as a separate instrument, sharing only the limb darkening coefficients. The ground-based follow-up photometry of TOI-3288 is detrended for airmass using a linear regression model. For the ExTrA data, we instead use a GP with a Matérn kernel to model correlated noise.

The orbital solutions from the joint fit with \texttt{Juliet} are presented in Table~\ref{table:orbital_solution}. The fitted parameters are directly obtained from the model, with reported uncertainties corresponding to the $1\sigma$ errors. Instrument-specific parameters, also derived from the model, are listed in App.~\ref{app:limb_instrument_table}. Derived parameters are computed from the fitted ones. In detail, the planetary radius $R_{\rm pl}$ is derived from the stellar radius and the planet-to-star radius ratio. The planetary mass $M_{\rm pl}$ is calculated using the RV equation, which relates the observed $K$ to the planetary mass. For transiting planets, $M_\mathrm{pl}\sin i \approx M_\mathrm{pl}$, since $i\approx 90^{\circ}$. The bulk density $\rho_{\rm pl}$ follows from the mass and radius, the semi-major axis $a_{\rm pl}$ is then computed via Kepler’s third law. The orbital inclination is estimated following the method of \citet{seager_unique_2003}, and the transit duration $T_{14}$ is derived using the method from \citet{kipping_characterizing_2014}. The planet’s equilibrium temperature $T_\mathrm{eq}$ is estimated from the stellar effective temperature, stellar radius, and semi-major axis, assuming a Bond albedo $A=0$. Finally, the incident stellar flux $S_{\rm pl}$ is computed from the stellar luminosity $L$ and the semi-major axis $a_{\rm pl}$.

\begin{table*}
\caption{Fitted and derived parameters for the giants presented in this paper.}
\label{table:orbital_solution}
\begin{center}
\begin{tabular}{l l l c c c c c c}
\hline\hline
\multicolumn{2}{l}{Parameter} &  & \multicolumn{3}{c}{TOI-3288 b} & \multicolumn{3}{c}{TOI-4666 b} \\ \hline
Fitted parameters & \\

\footnotesize{Orbital period} & $P$ & [days]  & \multicolumn{3}{c}{$1.4338635\pm 0.0000004$ } & \multicolumn{3}{c}{$2.9089168_{-0.0000006}^{+0.0000007}$ }\\
\footnotesize{Time of transit} & $T_0$ & [rBJD]$^{(a)}$  & \multicolumn{3}{c}{$59057.7310\pm 0.0002$ } & \multicolumn{3}{c}{$59168.4744\pm 0.0002$ }\\
\footnotesize{Radius ratio} & $R_\mathrm{pl}/R_{\star}$ &  & \multicolumn{3}{c}{$0.153\pm 0.001$ } & \multicolumn{3}{c}{$0.1934_{-0.0008}^{+0.0009}$ }\\
\footnotesize{Impact parameter} & $b$ &  & \multicolumn{3}{c}{$0.11_{-0.07}^{+0.09}$ } & \multicolumn{3}{c}{$0.09\pm 0.05$ }\\
\footnotesize{Stellar density} & $\rho_{\star}$ & [$\rho_\odot$]  & \multicolumn{3}{c}{$2.59_{-0.10}^{+0.07}$ } & \multicolumn{3}{c}{$3.11_{-0.05}^{+0.04}$ }\\
\footnotesize{Eccentricity} & $e$ (fixed) & & \multicolumn{3}{c}{0}& \multicolumn{3}{c}{0}\\
\footnotesize{RV semi-amplitude} & $K$ & [m s$^{-1}$]  & \multicolumn{3}{c}{$511\pm 19$ } & \multicolumn{3}{c}{$144_{-12}^{+11}$ }\\ \hline
Derived parameters & \\
\footnotesize{Planetary radius} & $R_\mathrm{pl}$ & [$R_\mathrm{Jup}$]  & \multicolumn{3}{c}{$1.00\pm 0.03$ } & \multicolumn{3}{c}{$1.11\pm 0.04$ }\\
\footnotesize{Planetary mass} & $M_\mathrm{pl}$ & [$M_\mathrm{Jup}$]  & \multicolumn{3}{c}{$2.11\pm 0.08$ } & \multicolumn{3}{c}{$0.70\pm 0.06$ } \\
\footnotesize{Planetary bulk density} & $\rho_\mathrm{pl}$ & [g cm$^{-3}$]  & \multicolumn{3}{c}{$2.6_{-0.2}^{+0.3}$ } & \multicolumn{3}{c}{$0.64_{-0.08}^{+0.09}$ } \\
\footnotesize{Inclination} & $i$ & [$^{\circ}$]  & \multicolumn{3}{c}{$89.1_{-0.8}^{+0.6}$ } & \multicolumn{3}{c}{$89.6\pm 0.2$ } \\
\footnotesize{Semi-major axis} & $a_\mathrm{pl}$ & [mAU]  & \multicolumn{3}{c}{$21.5\pm 0.1$ } & \multicolumn{3}{c}{$33.3\pm 0.2$ } \\
\footnotesize{Transit duration} & $T_{14}$ & [hours]  & \multicolumn{3}{c}{$1.93\pm 0.01$ } & \multicolumn{3}{c}{$2.370_{-0.008}^{+0.007}$ }\\
\footnotesize{Equilibrium temperature}$^{(b)}$ & $T_\mathrm{eq}$ & [K]  & \multicolumn{3}{c}{$1059\pm 21$ } & \multicolumn{3}{c}{$713\pm 14$ } \\ 
\footnotesize{Insolation} & $S_\mathrm{pl}$ & [$S_{\oplus}$]  & \multicolumn{3}{c}{$239\pm 7$ } & \multicolumn{3}{c}{$54\pm 3$ } \\
\hline
Spectroscopy parameters & &  & \footnotesize{NIRPS} & \multicolumn{2}{c}{\footnotesize{HARPS}} & \footnotesize{NIRPS} & \multicolumn{2}{c}{\footnotesize{HARPS}}\\
& &  & \footnotesize{HE} & \footnotesize{EGGS} & \footnotesize{HAM}  & \footnotesize{HE} & \footnotesize{EGGS} & \footnotesize{HAM} \\
Systemic RV & $\gamma$ & [km s$^{-1}$]  & \multicolumn{1}{c}{\begin{tabular}{@{}c@{}}$29.83$\\ $\pm 0.02$\end{tabular}} & \multicolumn{1}{c}{\begin{tabular}{@{}c@{}}$30.03$\\ $_{-0.03}^{+0.02}$\end{tabular}} & \multicolumn{1}{c}{\begin{tabular}{@{}c@{}}$29.73$\\ $\pm 0.03$\end{tabular}} & \multicolumn{1}{c}{\begin{tabular}{@{}c@{}}$7.47$\\ $\pm 0.01$\end{tabular}} & \multicolumn{1}{c}{\begin{tabular}{@{}c@{}}$8.02$\\ $\pm 0.01$\end{tabular}} & \multicolumn{1}{c}{\begin{tabular}{@{}c@{}}$7.61$\\ $\pm 0.03$\end{tabular}}\\
Jitter & $\sigma$ & [m s$^{-1}$]  & \multicolumn{1}{c}{$80_{-15}^{+13}$ } & \multicolumn{1}{c}{$1.0_{-1.0}^{+33.7}$ } & \multicolumn{1}{c}{$0.2_{-0.2}^{+8.4}$ } & \multicolumn{1}{c}{$22_{-21}^{+26}$ } & \multicolumn{1}{c}{$0.1_{-0.1}^{+5.2}$ } & \multicolumn{1}{c}{$0.7_{-0.7}^{+22.9}$ }\\
Residual noise & $\mathrm{RMS}\left(\mbox{O-C}\right)$ & [m s$^{-1}$]  & 92 & 94 & 41 & 70 & 89 & 79 \\ \hline
\end{tabular}
\end{center}
\textbf{Notes:} The limb darkening and photometric instrumental parameters can be found in Appendix \ref{app:limb_instrument_table}.\\
The values assumed for the solar and planetary constants are the IAU 2015 Resolution B 3 values from \citep{prsa_nominal_2016}.\\
$^{(a)}$ The reduced Barycentric Julian Date in Barycentric Dynamical Time ($\mathrm{rBJD_{TDB}}$), obtained by subtracting 2 400 000 from the $\mathrm{BJD_{TDB}}$.\\
$^{(b)}$ Assuming a Bond albedo $A=0$.\\
\end{table*}

\subsection{TOI-3288}
\begin{figure}
    \includegraphics[width=\linewidth]{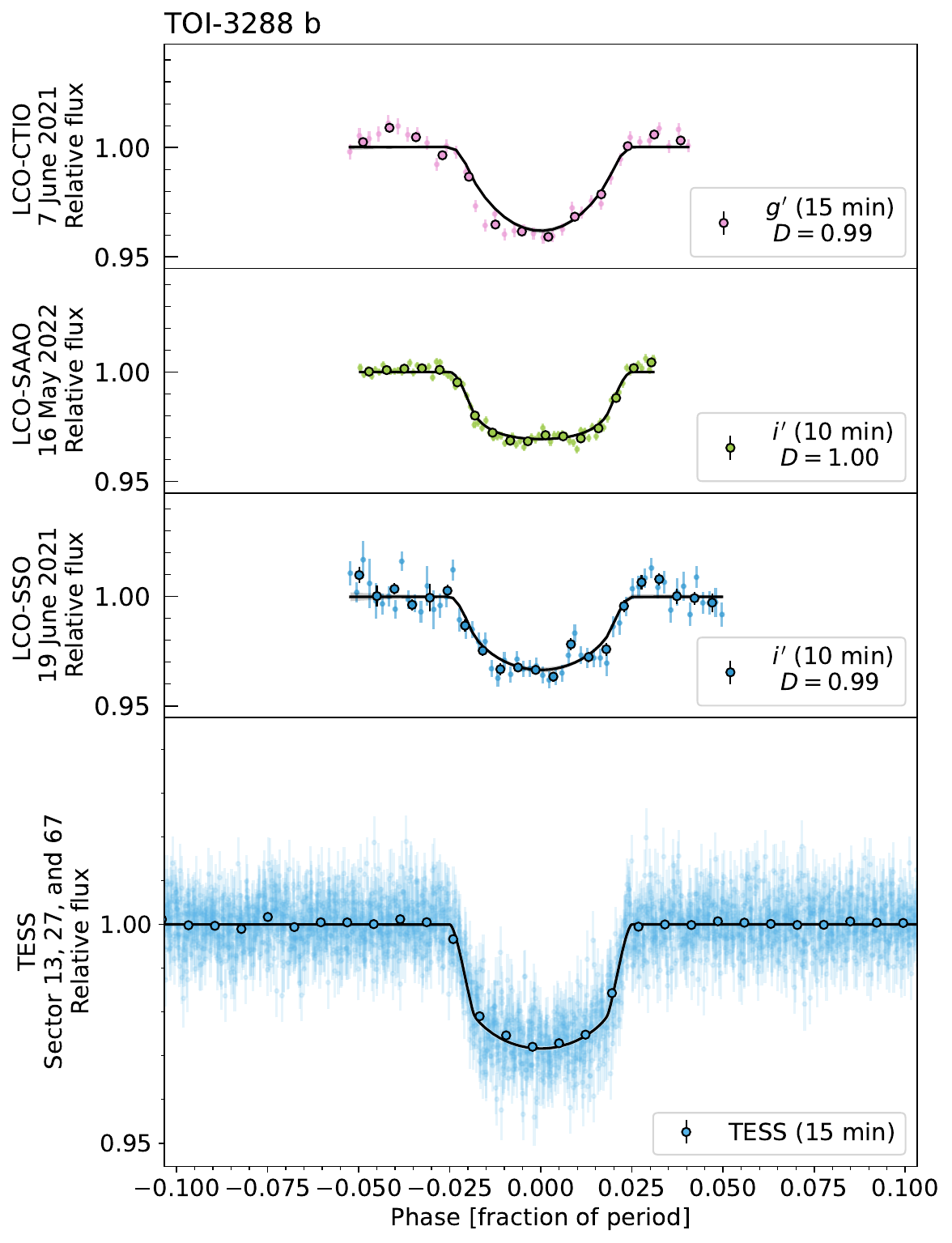}
    \caption{Phase-folded light curves of TOI-3288 from \textit{TESS}-SPOC, QLP, and LCO (SSO, SAAO, and CTIO). If a dilution factor was fitted, it is indicated in the legend of each subpanel. Markers with black edges represent data binned per 10 or 15 minutes, as noted in the legend.}
    \label{fig:TOI-3288_SG1}
\end{figure}

TOI-3288 b is a $2.11 \pm 0.08\mjup$ gas giant planet transiting a K9V star every 1.4 days. With a radius of $1.00\pm0.03\rjup$ and a equilibrium temperature of $1059\pm20~\mathrm{K}$ it is a relatively massive Hot Jupiter. Fig.~\ref{fig:TOI-3288_SG1} shows the photometric data with the derived median posterior model. 

\begin{figure}
    \centering
    \includegraphics[width=\linewidth]{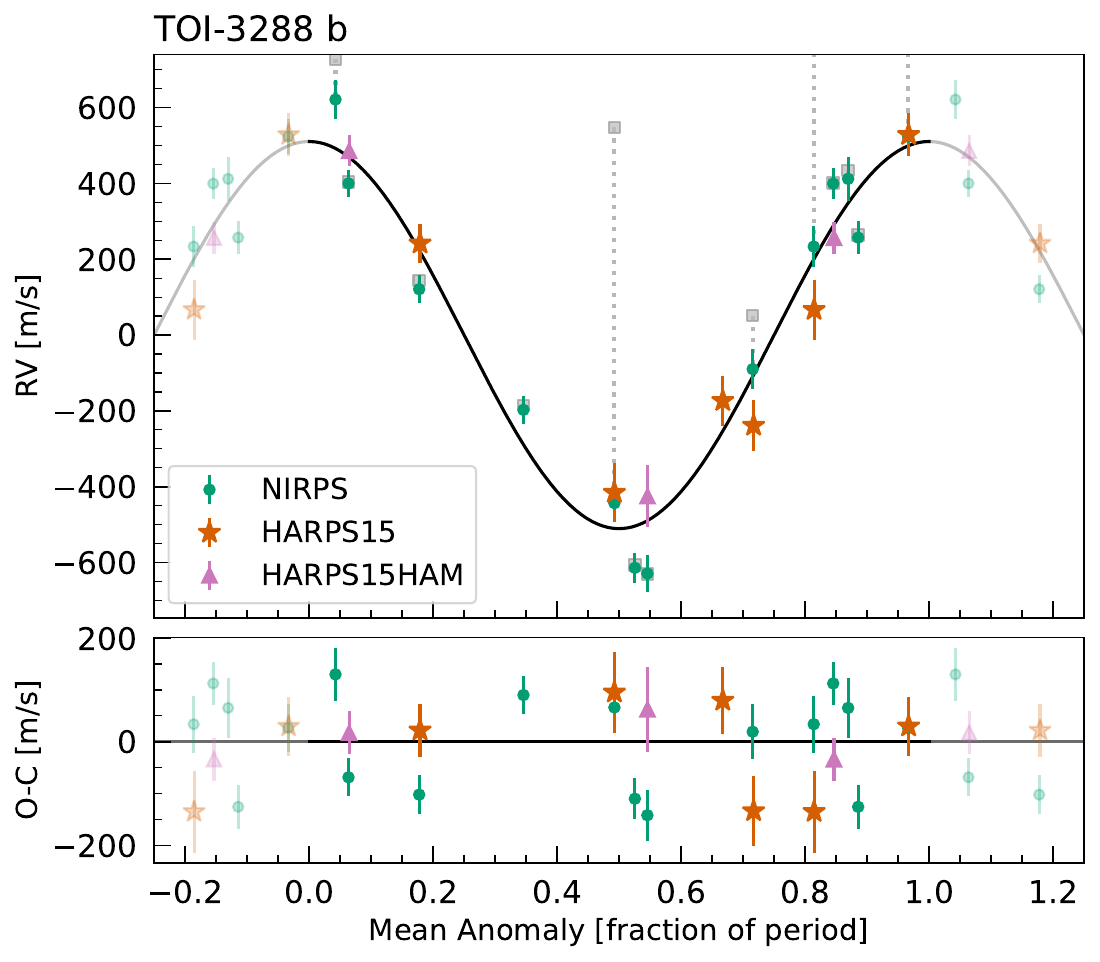}
    \caption{Night-binned RVs of TOI-3288 with the \texttt{Juliet} orbital solution overplotted as a black line. The residuals have an RMS of $\sim 65\,\mathrm{m\,s^{-1}}$. The RVs have been cleaned on telluric contamination. The RVs from before the post-processing steps are shown as gray squares, with dotted lines indicating the difference from the corrected values.}
    \label{fig:TOI3288_RV}
\end{figure}

Fig. \ref{fig:TOI3288_RV} shows the post-processed HARPS and NIRPS RVs. Separately, HARPS finds $K = 486\pm 33\,\mathrm{m\,s^{-1}}$, and NIRPS finds $K = 532\pm33~\mathrm{m\,s^{-1}}$, consistent within $1\,\sigma$. Combined, the two instruments give $K = 511 \pm 19~{\rm m\,s^{-1}}$. Table~\ref{table:orbital_solution} presents the joint fit results.

\subsection{TOI-4666}
TOI-4666 is a M2.5 main-sequence star hosting a $0.70\pm0.06\mjup$ giant planet. Despite its short orbital period of 2.9 days, the planet's equilibrium temperature is only $713\pm14~\mathrm{K}$ due to the low insolation flux ($54\,\pm 3~S_\oplus$) received from its cool host, placing it outside the typical Hot Jupiter range.

\begin{figure*}
    \includegraphics[width=\linewidth]{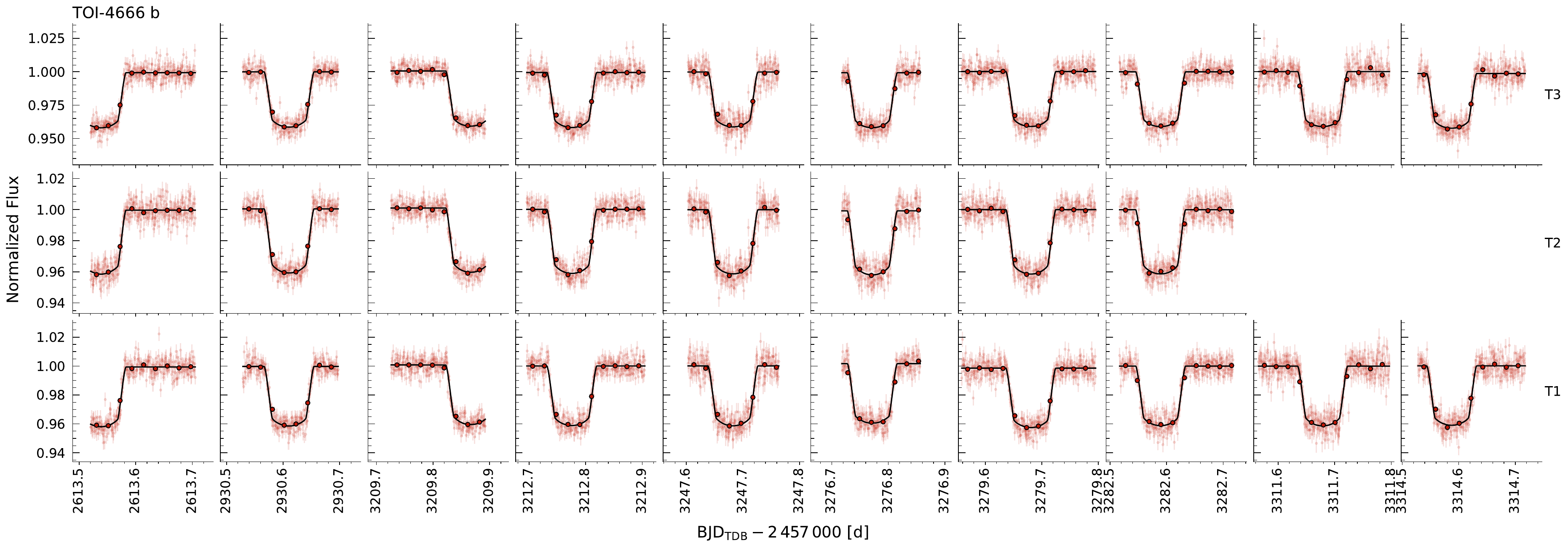}
    \caption{The normalized flux versus time of TOI-4666 transits observed with ExTrA, shown per telescope. The \texttt{Juliet} fit is represented by the black line, and the black markers with black outlines represent data binned in 30-minute intervals.}
    \label{fig:TOI4666_ExTrA}
\end{figure*}

\begin{figure}
    \includegraphics[width=\linewidth]{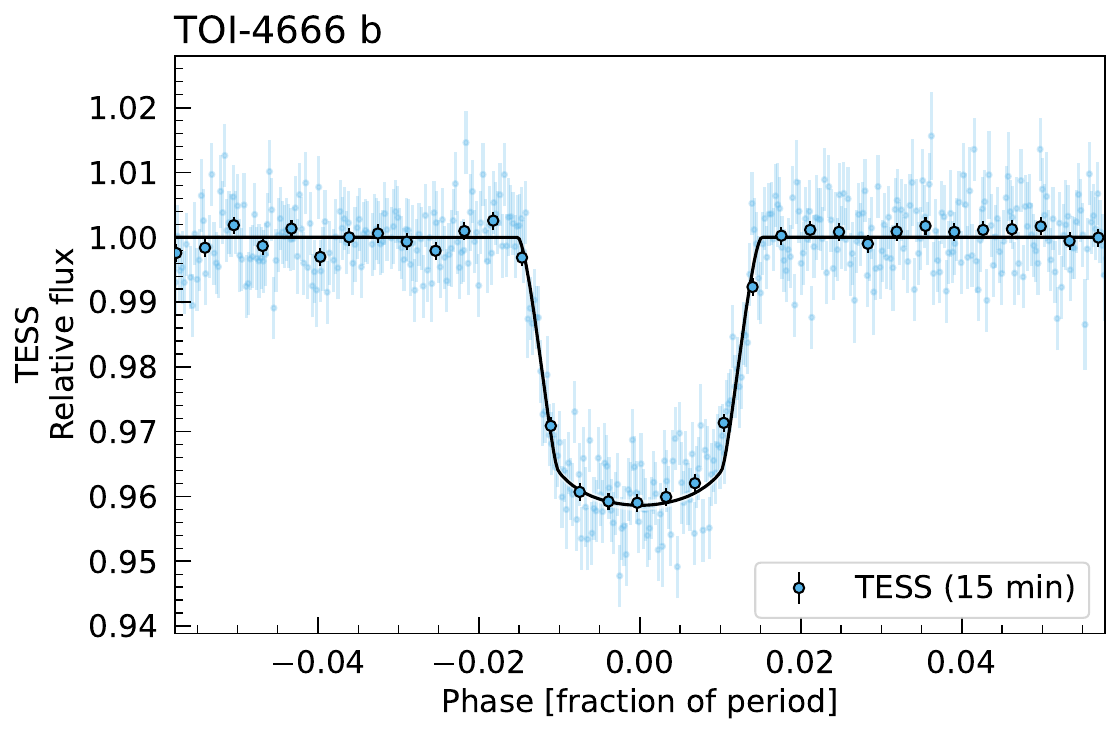}
    \caption{Phase-folded \textit{TESS} light curve of TOI-4666, with the \texttt{Juliet} fit shown as a black line. Markers with black edges indicate 15-minute binned data.}
    \label{fig:TOI4666_TESS}
\end{figure}

Fig.~\ref{fig:TOI4666_ExTrA}, and \ref{fig:TOI4666_TESS}, show the photometry data with the median posterior model. The orbital period is refined with greater precision by the additional transits observed by ExTrA. The ExTrA data show residual instrumental systematics, as the linear regression airmass correction does not fully account for the deviations from the out-of-transit baseline. However, no GP was included in the joint fit to avoid computational cost. Each transit and telescope is treated as a separate instrument with distinct systematics, except for the limb-darkening coefficient, which is shared for all three telescopes. 


\begin{figure}
    \includegraphics[width=\linewidth]{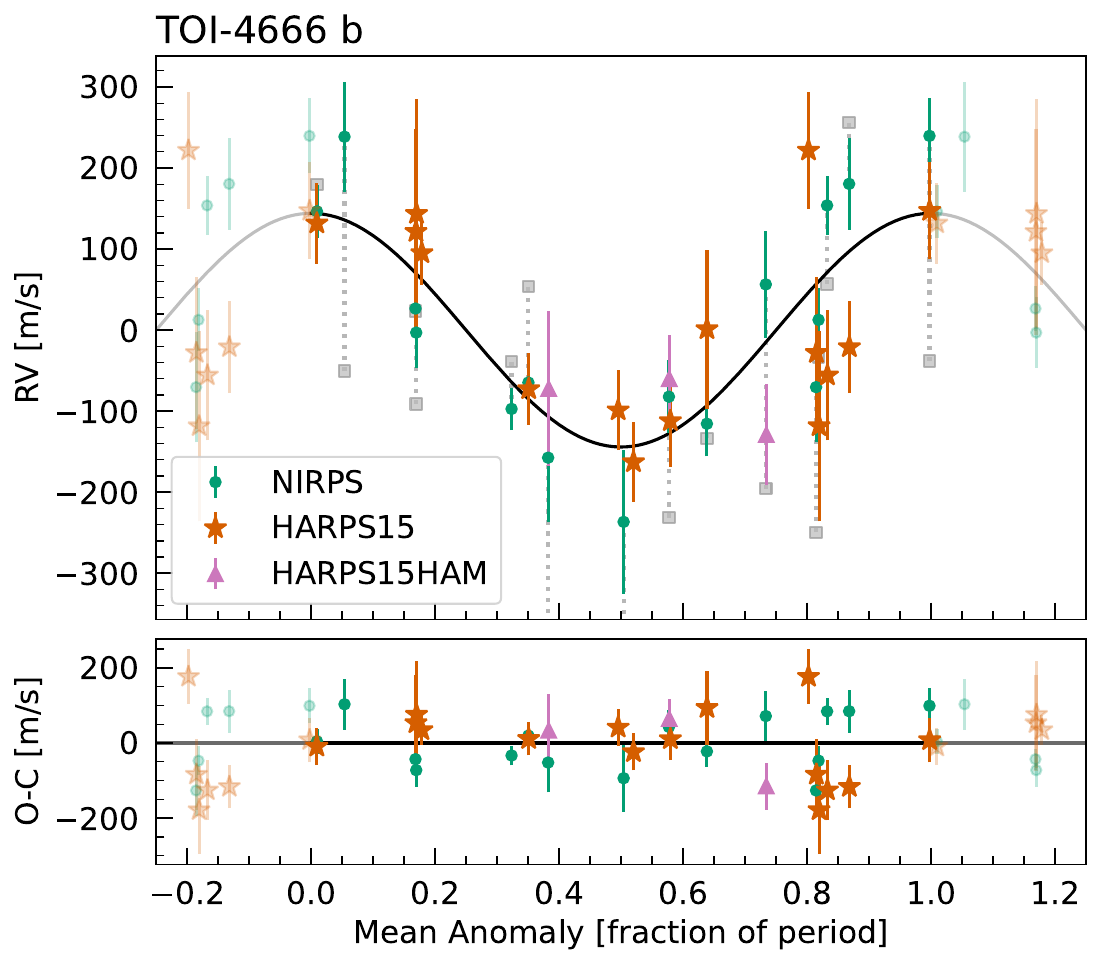}
    \caption{Night-binned RVs of TOI-4666 with the \texttt{Juliet} orbital solution overplotted. The residuals have an RMS of $\sim75\,\mathrm{m\,s^{-1}}$. The RVs before the post-processing steps are shown as gray squares, with dotted lines indicating the difference from the corrected values.}
    \label{fig:TOI4666_RV}
\end{figure}

Fig.~\ref{fig:TOI4666_RV} shows the HARPS and NIRPS RVs. When fitted separately, HARPS finds $K = 114\pm22\,\mathrm{m\,s^{-1}}$ and NIRPS $K=162\pm21\,\mathrm{m\,s^{-1}}$, consistent within $1.6\,\sigma$, indicating relatively good agreement between the optical and nIR. The joint fit results in $K = 144^{+11}_{-12}~\mathrm{m\,s^{-1}}$.

\subsection{Eccentricity}
In our analysis, we adopt fixed values for the orbital eccentricity and argument of periastron ($e=0$ and $\omega=90^{\circ}$). Allowing these parameters to vary yields $1\,\sigma$ upper limits on the eccentricity of $0.03$ for TOI-3288 b and $0.10$ for TOI-4666 b, while $\omega$ remains unconstrained. Model comparison using the log-evidence from our \texttt{Juliet} fits indicate that both planets are best described by $e=0$ solutions. Specifically, when fitting only the RV data, the log-evidence for TOI-3288 is $-11.118\pm 0.435$ assuming a circular orbit, and $-13.930 \pm 0.456$ including eccentricity. For TOI-4666 a circular orbit has a log-evidence of $6.218 \pm 0.438$, decreasing to $4.585 \pm 0.431$ with eccentricity. This is consistent with close-in giants ($P \lesssim 3\,\mathrm{d}$) rapidly circularizing due to strong tidal dissipation \citep{hut_tidal_1981}.

\section{Discussion}
\label{sec:Discussion}
\begin{figure*}
    \centering
    \includegraphics[width=\linewidth]{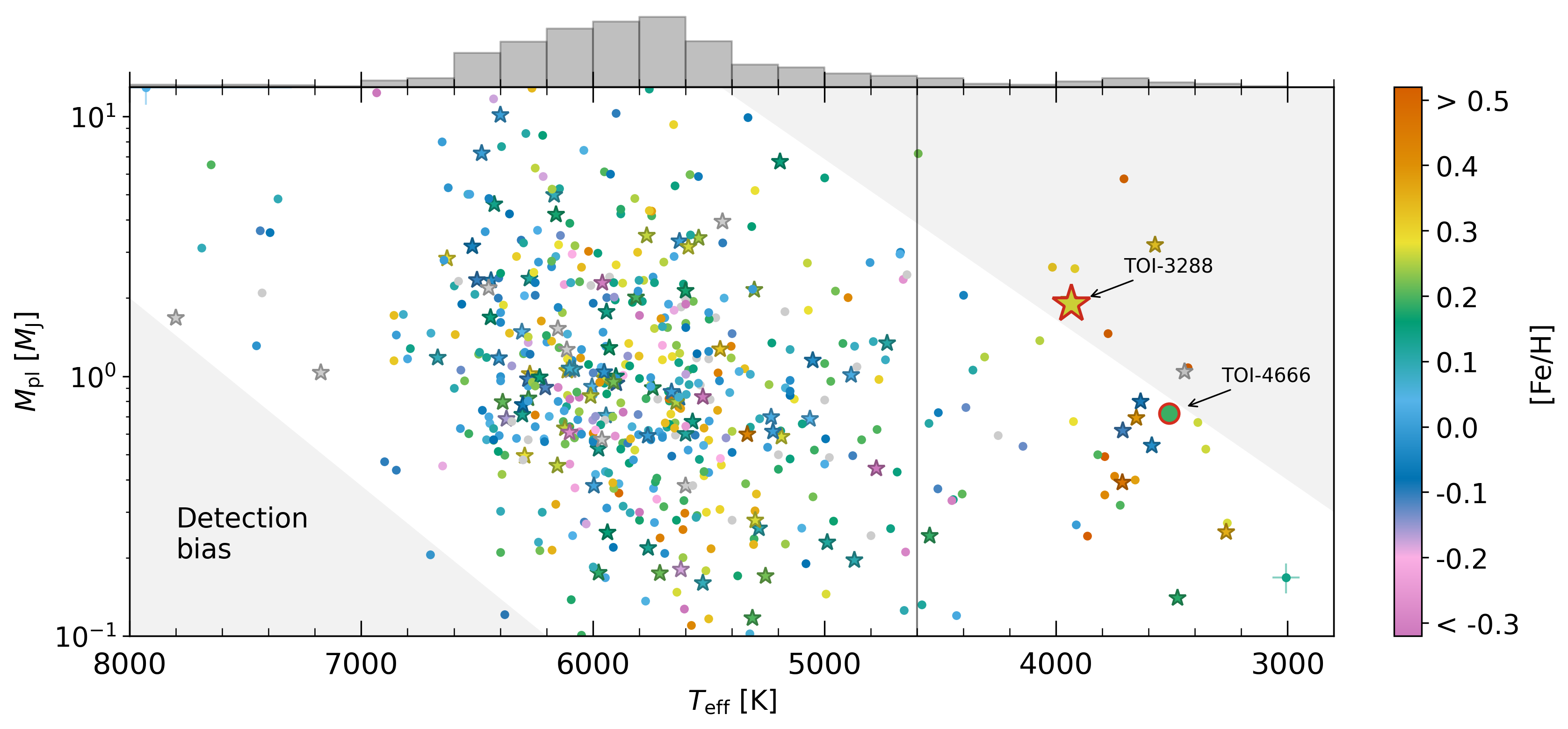}
    \caption{TOI-3288 b and TOI-4666 b (red encircled) compared to known transiting exoplanets from the PlanetS catalog \citep[][updated August 2025]{otegi_revisited_2020, parc_super-earths_2024}. 
    Marker colours show stellar metallicity, with gray for stars of unknown metallicity. Stellar markers indicate planetary companions in binary systems, circles are single stars. The gray vertical line marks $4600$~K. The relative occurrence of the gas giants with masses between 0.1 and $13\mjup$ is shown as a histogram at the top. Our two mass characterizations contribute to a rare population of exoplanets.}
    \label{fig:overview}
\end{figure*}

With NIRPS, we characterize two gas giants orbiting low-mass stars: TOI-3288 b and TOI-4666 b. As shown in Fig.~\ref{fig:overview}, these companions occupy a relatively sparse region of the known transiting exoplanets. Through the NIRPS-GTO subprogram, we are helping to fill in the exoplanet population around low-mass stars, using NIRPS’s ability to observe faint M dwarfs. In the following subsections, we discuss the age, interior structure, binarity, metallicity, and atmospheric properties of these planets, as well as the effect of the remnant telluric contamination correction on the RVs.

\subsection{Age}
If the periodicity detected in the \textit{TESS} light curves is not due to instrumental artifacts, the implied stellar rotation periods are relatively short for these type of stars \citep{barnes_rotational_2003, barnes_ages_2007}. Such short rotation periods could indicate they are young. TOI-3288 has a 24~\% probability of being associated with the tidal tail of UPK612, which has an estimated age of $\sim100$ Myr \citep{kos_tidal_2024}. TOI-4666, however, is not known to be associated with any cluster. Using the stellar parameters ($T_{\rm eff}$, [Fe/H], $\log g$, $L_{\rm bol}$, and $\omega$; see Tab.\ref{table:stellar_parameters}), we estimated ages from PARSEC isochrones \citep{bressan_span_2012} via the PARAM1.5 web interface\footnote{\url{https://stev.oapd.inaf.it/cgi-bin/param}}. We find ages of $590^{+300}_{-200}$ Myr for TOI-3288 and $80\pm6$ Myr for TOI-4666. 
Since TOI-4666 is not affiliated with a young cluster, there is no other strong evidence that it is this young. 

\subsection{Interior}
We used the planetary evolution code \texttt{completo} \citep{mordasini_characterization_2012} to estimate the heavy element mass of both planets. The planet is modelled with a core of iron and silicates with a mass up to $\rm 10\,M_{\oplus}$, an envelope along with a semi-grey atmospheric model. The envelope contains hydrogen and helium modelled with the equation of state (EoS) from \cite{chabrier_new_2021}, and the heavy elements are modelled as water with the EoS AQUA from \cite{haldemann_aqua_2020}. According to the empirical formulas of \citet{sestovic_investigating_2018}, TOI-3288 b and TOI-4666 b are not inflated, as both do not exceed the threshold in incident flux for their planetary masses. Given this relatively low irradiation, no inflation parameters are included in the modelling.

We built a grid of evolution models coupled with a Bayesian inference scheme to retrieve the heavy element content compatible with the observed planetary parameters. For TOI-3288 b, we used a normal prior on the stellar age centred on 0.6 Gyrs. We find a heavy element fraction in the envelope of $\rm Z = 0.28\pm 0.04$ and a total heavy element mass of $\rm 190\pm 25~M_\oplus$. For TOI-4666, using a normal prior around 300 Myr, we find a heavy element fraction in the envelope of $\rm Z = 0.12\pm 0.06$ and a total heavy element mass of $\rm 31\pm 14~M_\oplus$. We note that at young ages the interior models may be sensitive to the initial conditions and that different interior models may lead to variations of the derived heavy element content.

\subsection{Binarity}
\label{sec:Binarity}
Many of the lower-mass giants around low-mass stars are in binaries, hinting that, besides metallicity, binarity might also enhance giant planet formation. In Fig.~\ref{fig:overview} stellar-shaped markers indicate companions in binary systems. We examined their distribution to assess whether binarity affects formation, and found that 17~\% of stars hosting giant planets with $T_{\rm eff} \geq 4600~{\rm K}$ are in binary systems, compared to 24~\% for $T_{\rm eff} < 4600~{\rm K}$ and 39~\% for $T_{\rm eff} < 3890~{\rm K}$.

TOI-3288 is part of a wide binary system with an angular separation of approximately $2.3\arcsec$, corresponding to a physical separation of $\sim 435~{\rm AU}$. The presence of a gravitationally bound companion is confirmed using astrometric data from \textit{Gaia} DR3, which reveals consistent parallaxes and proper motions between the two stars \citep{mugrauer_gaia_2023}. Assuming the two components have the same age and are both in the main sequence, the G-band magnitude of 17.2 mag for the stellar companion translates into a spectral type close to M4V according to the tabulated relations of \cite{pecaut_intrinsic_2013}.
 
The formation of giant planets around low-mass stars presents a challenge to standard core accretion models, which predict limited planetary growth due to the reduced mass reservoir in their protoplanetary disk. \cite{fabrycky_shrinking_2007} have shown that wide binary systems $(>100~{\rm AU})$ may facilitate planet migration through mechanisms such as Kozai-Lidov cycles and dynamical interactions. Alternatively, simulations suggest that stellar companions at separations of a few hundred AU can trigger gravitational disk instability, leading to the formation of massive planets and brown dwarfs \citep{cadman_binary_2022}.
 
Observational studies have shown the most massive short-period giant planets are preferentially found in multiple star systems \citep{zucker_mass-period_2002, eggenberger_detection_2004, desidera_properties_2007, fontanive_census_2021}. \cite{ngo_friends_2016} have shown that the occurrence rate of hot Jupiters is 2.9 times larger for binaries with separation in the range 50-2000 AU than for field stars. Similarly, \cite{eeles-nolle_high_2025} found elevated stellar multiplicity rates for hot Jupiter hosts. 
 
The presence of a stellar companion may have altered the formation, migration and evolution of TOI-3288 b, through gravitational perturbations affecting the protoplanetary disk. Given its physical separation ($<1000~{\rm AU}$), the stellar companion to TOI-3288 falls within the regime where migration and dynamical interactions may be at play. While the companion is unlikely to have truncated the protoplanetary disk, it may have influenced the formation and evolution of the planet and contributed to orbital evolution processes.

\subsection{Metallicity}
\label{sec:Metallicity}
Metal-rich stars are more likely to host giant exoplanets, as described by the planet-metallicity correlation for FGK stars \citep{ida_toward_2004, santos_spectroscopic_2004, fischer_planetmetallicity_2005} and for M dwarfs \citep{gan_metallicity_2025}. Here, we report an additional tentative trend in the observations: more massive gas giants are hosted by more metal-rich stars.

There are regions in Fig.~\ref{fig:overview} where few giant planets are found, which we have defined by eye and shaded in gray to emphasize the observed scarcity. The shaded area for $T_{\rm eff} \gtrsim 6000~{\rm K}$ reflects detection biases, while the second region highlights the apparent upper envelope of massive gas giants around low-mass stars. Although low-mass giants are challenging to characterize around low-mass stars, massive gas giants should be easily detectable. Their paucity therefore suggests that such planets rarely form around low-mass stars.

When comparing the giants around low-mass stars ($T_{\rm eff} < 4600$ K) inside this region to those outside, a tentative trend emerges: the giants within the scarcely populated region tend to orbit more metal-rich host stars (mean metallicity $\sim 0.44$ dex) than those outside ($\sim0.16$ dex). This aligns with theoretical expectations \citep{thommes_gas_2008, mordasini_extrasolar_2012} where low-mass stars are thought to host less massive protoplanetary disks, but high metallicity can compensate for lower disk mass \citep{vorobyov_mass_2008, alibert_extrasolar_2011, hobson_sophie_2018}, and vice versa. 

\subsection{NIRPS' faint end}
On the faint end of NIRPS we are more limited by telluric contamination than on the bright side. The technique proposed in Sec. \ref{sec:telluric_contamination}, reduces remnant contamination. 
Using an RV-only \texttt{Juliet} fit to the uncorrected and corrected RVs, the RMS of the residuals for TOI-3288 decreases from $373~\mathrm{m\,s^{-1}}$ to $92~\mathrm{m\,s^{-1}}$, and for TOI-4666 from $135~\mathrm{m\,s^{-1}}$ to $70~\mathrm{m\,s^{-1}}$. 
With this relatively straightforward approach, we can ensure that no telluric contamination influences the remaining RVs. Any remaining scatter is introduced by the low S/N of our observations. 
The remaining RMS is slightly larger than the median photon noise of the observations, which could suggest underestimated errors. This underestimation is not due to the BERV-window exclusion. While excluding this region reduces information and thus increases the errors, the OH-dominated region was already introducing noise, so the effect is not expected to be significant. 
Instead, the excess RMS could stem from stellar activity, as expected for K9V and M2.5V hosts, especially if the estimated rotation periods and ages are reliable. We do not attribute this remaining signal to an additional planet, as the S/N in both the NIRPS and HARPS data is too low to draw conclusions.

\subsection{Atmospheric characterization}
The transiting nature and short orbital periods of these planets make them possible targets for atmospheric characterization. We assess their feasibility using the transmission and emission spectroscopy metrics (TSM and ESM) defined by \citet{kempton_framework_2018}. Both planets show potential for atmospheric follow-up, although they may be less favourable than some well-known hot Jupiters. TOI-3288 b has an ESM of 104, a TSM of 30, and a scale height of 117 km. TOI-4666 b has an ESM of 77, a TSM of 109, and a scale height of 309 km. The latter may be particularly interesting, as it orbits an early M dwarf and is classified as a warm Jupiter. However, their magnitudes ($V\sim15$, $J\sim12$, $H\sim11$) may limit the availability of facilities for high-resolution spectroscopic follow-up. While warm Jupiters are typically harder to study, the ELT will enable this. The upcoming ELT spectrographs ANDES and METIS aim to characterize smaller and warmer planets than the typical ultra-hot Jupiters. As giant planets around M dwarfs are rare, and only a handful have favourable conditions for atmospheric follow-up, TOI-3288 b and TOI-4666 b stand out as valuable opportunities to probe giant-planet atmospheres in the low-mass-star regime.

\section{Conclusions}
\label{sec:Conclusions}
In this work we confirm and characterize new gas giants transiting low-mass stars identified by \textit{TESS}. We post-processed the NIRPS DRS spectra to further mitigate telluric contamination, excluding a window centred on the BERV from the Gaussian fit when determining RVs from a CCF. From these cleaned RVs, we find that TOI-3288 b is a dense hot Jupiter in a system with a stellar companion, while TOI-4666 b is a low-density warm Jupiter, likely with a puffy atmosphere. Their unique properties make them valuable targets for exploring giant planet formation around low-mass stars, where binarity and metallicity may play key roles. The NIRPS-GTO subprogram will continue to characterize transiting gas giants, in parallel with CORALIE which focuses on late K dwarfs \citep{frensch_three_2025}, to further explore the apparent decline of massive gas giants around low-mass stars.

\section*{Data availability}
Tables \ref{tab:TOI-3288_RVs}, \ref{tab:TOI-4666_RVs}, \ref{tab:TOI-3288_phot} and \ref{tab:TOI-4666_phot} are available in electronic form at the CDS via anonymous ftp to \url{cdsarc.u-strasbg.fr (130.79.128.5)} or via \url{http://cdsweb.u-strasbg.fr/cgi-bin/qcat?J/A+A/}.

\bibliographystyle{aa}
\bibliography{references.bib}

\begin{acknowledgements}
Some of the observations in this paper made use of the High-Resolution Imaging instrument Zorro and were obtained under Gemini LLP Proposal Number: GN/S-2021A-LP-105. Zorro was funded by the NASA Exoplanet Exploration Program and built at the NASA Ames Research Center by Steve B. Howell, Nic Scott, Elliott P. Horch, and Emmett Quigley. Zorro was mounted on the Gemini South telescope of the international Gemini Observatory, a program of NSF’s OIR Lab, which is managed by the Association of Universities for Research in Astronomy (AURA) under a cooperative agreement with the National Science Foundation. on behalf of the Gemini partnership: the National Science Foundation (United States), National Research Council (Canada), Agencia Nacional de Investigaci\'on y Desarrollo (Chile), Ministerio de Ciencia, Tecnolog\'ia e Innovaci\'on (Argentina), Minist\'erio da Ci\^encia, Tecnologia, Inova\c{c}\~oes e Comunica\c{c}\~oes (Brazil), and Korea Astronomy and Space Science Institute (Republic of Korea).\\
AL  acknowledges support from the Fonds de recherche du Qu\'ebec (FRQ) - Secteur Nature et technologies under file \#349961.\\
AL, LMa, RA, BB, CC, RD, \'EA \& FBa  acknowledge the financial support of the FRQ-NT through the Centre de recherche en astrophysique du Qu\'ebec as well as the support from the Trottier Family Foundation and the Trottier Institute for Research on Exoplanets.\\
The Board of Observational and Instrumental Astronomy (NAOS) at the Federal University of Rio Grande do Norte's research activities are supported by continuous grants from the Brazilian funding agency CNPq. This study was partially funded by the Coordena\c{c}\~ao de Aperfei\c{c}oamento de Pessoal de N\'ivel Superior—Brasil (CAPES) — Finance Code 001 and the CAPES-Print program.\\
0\\
XDu  acknowledges the support from the European Research Council (ERC) under the European Union’s Horizon 2020 research and innovation programme (grant agreement SCORE No 851555) and from the Swiss National Science Foundation under the grant SPECTRE (No 200021\_215200).\\
This work has been carried out within the framework of the NCCR PlanetS supported by the Swiss National Science Foundation under grants 51NF40\_182901 and 51NF40\_205606.\\
LMa, RD, \'EA \& FBa  acknowledges support from Canada Foundation for Innovation (CFI) program, the Universit\'e de Montr\'eal and Universit\'e Laval, the Canada Economic Development (CED) program and the Ministere of Economy, Innovation and Energy (MEIE).\\
XB, XDe \& TF  acknowledge funding from the French ANR under contract number ANR\-24\-CE49\-3397 (ORVET), and the French National Research Agency in the framework of the Investissements d'Avenir program (ANR-15-IDEX-02), through the funding of the ``Origin of Life" project of the Grenoble-Alpes University.\\
ED-M, JGd, NCS, SCB \& EC  acknowledge the support from FCT - Funda\c{c}\~ao para a Ci\^encia e a Tecnologia through national funds by these grants: UIDB/04434/2020, UIDP/04434/2020.\\
ED-M  further acknowledges the support from FCT through Stimulus FCT contract 2021.01294.CEECIND. ED-M  acknowledges the support by the Ram\'on y Cajal contract RyC2022-035854-I funded by MICIU/AEI/10.13039/501100011033 and by ESF+.\\
NN, JIGH, ASM \& RR  acknowledge financial support from the Spanish Ministry of Science, Innovation and Universities (MICIU) projects PID2020-117493GB-I00 and PID2023-149982NB-I00.\\
NN  acknowledges financial support by Light Bridges S.L, Las Palmas de Gran Canaria.\\
NN acknowledges funding from Light Bridges for the Doctoral Thesis "Habitable Earth-like planets with ESPRESSO and NIRPS", in cooperation with the Instituto de Astrof\'isica de Canarias, and the use of Indefeasible Computer Rights (ICR) being commissioned at the ASTRO POC project in the Island of Tenerife, Canary Islands (Spain). The ICR-ASTRONOMY used for his research was provided by Light Bridges in cooperation with Hewlett Packard Enterprise (HPE).\\
KAM  acknowledges support from the Swiss National Science Foundation (SNSF) under the Postdoc Mobility grant P500PT\_230225.\\
RA  acknowledges the Swiss National Science Foundation (SNSF) support under the Post-Doc Mobility grant P500PT\_222212 and the support of the Institut Trottier de Recherche sur les Exoplan\`etes (IREx).\\
We acknowledge funding from the European Research Council under the ERC Grant Agreement n. 337591-ExTrA.\\
JRM  acknowledges CNPq research fellowships (Grant No. 308928/2019-9).\\
Co-funded by the European Union (ERC, FIERCE, 101052347). Views and opinions expressed are however those of the author(s) only and do not necessarily reflect those of the European Union or the European Research Council. Neither the European Union nor the granting authority can be held responsible for them.\\
GAW is supported by a Discovery Grant from the Natural Sciences and Engineering Research Council (NSERC) of Canada.\\
SCB   acknowledges the support from Funda\c{c}\~ao para a Ci\^encia e Tecnologia (FCT) in the form of a work contract through the Scientific Employment Incentive program with reference 2023.06687.CEECIND and DOI \href{https://doi.org/10.54499/2023.06687.CEECIND/CP2839/CT0002}{10.54499/2023.06687.CEECIND/CP2839/CT0002.}\\
NBC  acknowledges support from an NSERC Discovery Grant, a Canada Research Chair, and an Arthur B. McDonald Fellowship, and thanks the Trottier Space Institute for its financial support and dynamic intellectual environment.\\
DE  acknowledge support from the Swiss National Science Foundation for project 200021\_200726. The authors acknowledge the financial support of the SNSF.\\
ICL  acknowledges CNPq research fellowships (Grant No. 313103/2022-4).\\
BLCM  acknowledge CAPES postdoctoral fellowships.\\
BLCM  acknowledges CNPq research fellowships (Grant No. 305804/2022-7).\\
CMo  acknowledges the funding from the Swiss National Science Foundation under grant 200021\_204847 “PlanetsInTime”.\\
CP  acknowledges support from the NSERC Vanier scholarship, and the Trottier Family Foundation. CP  also acknowledges support from the E. Margaret Burbidge Prize Postdoctoral Fellowship from the Brinson Foundation.
\end{acknowledgements}

\appendix
\section{False Positives}
\label{sec:False_Positives}
\subsection{TOI-2341 (SB2)}
\begin{figure}[H]
    \includegraphics[width=\linewidth]{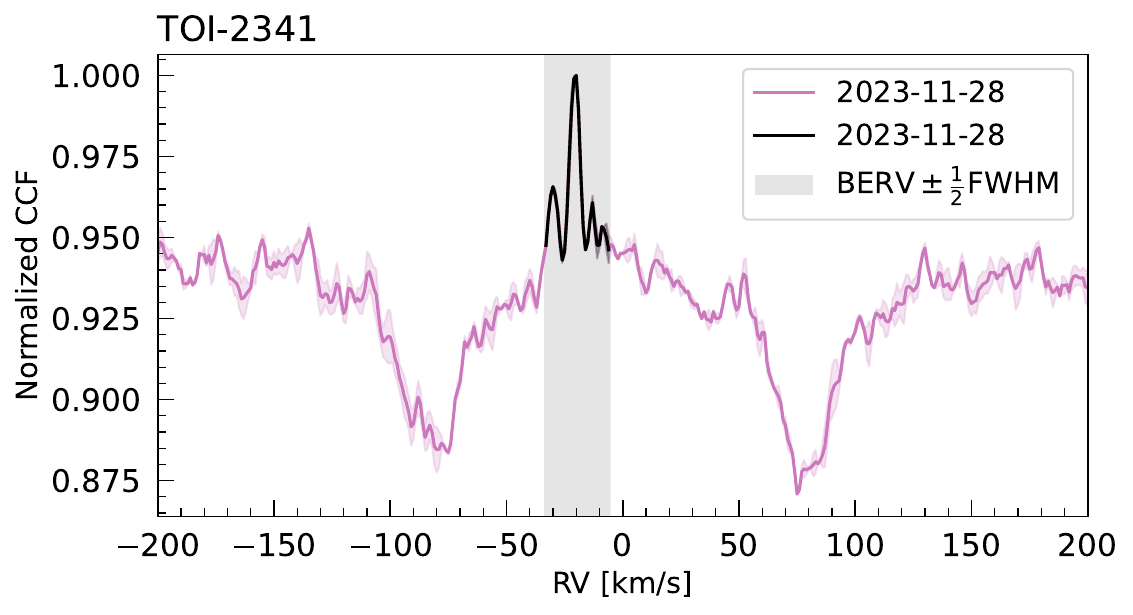}
    \caption{The normalized NIRPS CCF of the SB2 TOI-2341. The BERV is indicated by a gray shaded area, with limits defined by the FWHM of the stellar components in the CCF.}
    \label{fig:TOI-2341}
\end{figure}

TOI-2341 was early on recognized as a spectroscopic binary and was therefore observed only once. The two components in the CCF are visible in Fig.~\ref{fig:TOI-2341}.
\subsection{TOI-5295 (Triple system)}
\begin{figure}[H]
    \includegraphics[width=\linewidth]{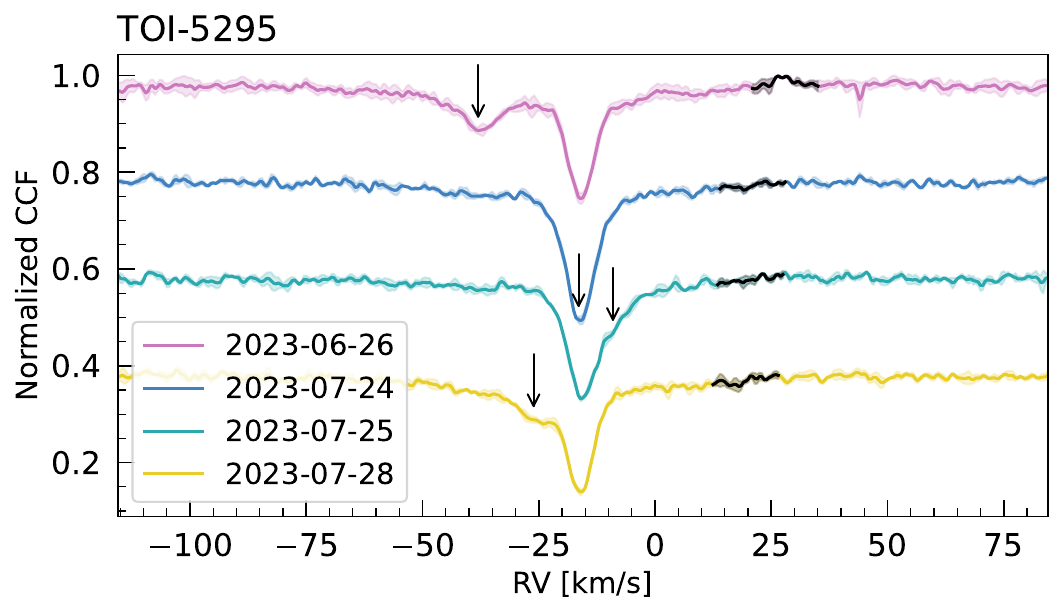}
    \caption{The normalized NIRPS CCFs of TOI-5295, where the RV variation of the second component is in agreement with the \textit{TESS} ephemerides. An arbitrary offset has been applied to the NIRPS CCFs to improve visibility. The BERV affected region is indicated in black.}
    \label{fig:TOI-5295}
\end{figure}

TOI-5295 shows a moving component in the NIRPS CCFs computed with an M1 mask (Fig.~\ref{fig:TOI-5295}). The RVs are derived by subtracting the average spectrum from each individual spectrum to suppress the primary signal, and then fitting a Gaussian with \texttt{iCCF} to the residual CCFs. After correcting for the movement of the primary component (negligible at 400~m~s$^{-1}$ compared to 40~km~s$^{-1}$), the RVs of the secondary are consistent with the \textit{TESS} ephemerides. A simple cosine fit results in a semi-amplitude of $K \sim 20$~km~s$^{-1}$. The secondary mass is estimated at $\sim0.4~M_\odot$, using the mass-luminosity relation of \citet{cuntz_mass-luminosity_2018}, the primary mass of $0.57~M_\odot$ from ExoFOP\footref{foot:EXOFOP}, and the contrast ratio of the primary and secondary component in the CCFs. The transit of TOI-5295 B is caused by a low-mass ($\sim95\mjup$) stellar companion, just above the hydrogen-burning limit, identifying this as a triple star system. 

\section{Fast rotators}
\label{app:Fast_Rotators}
\subsection{TOI-1227}
The young star TOI-1227 (11 Myr) was validated as a planet host by \citet{mann_TESS_2022} using SOAR \citep{tokovinin_ten_2018} and LCO photometry, resulting in a planetary radius of $0.85\rjup$. As this value falls within our program’s selection criteria, we attempted NIRPS follow-up. However, the star’s rapid rotation \citep[$1.65 \pm 0.04~{\rm d}$;][]{mann_TESS_2022} produces significantly broadened CCFs (${\rm FWHM} \sim 27~{\rm km~s^{-1}}$), and all observations coincided with a BERV crossing, preventing a reliable RV determination. We therefore stopped observations of this target.

\begin{figure}[H]
    \includegraphics[width=\linewidth]{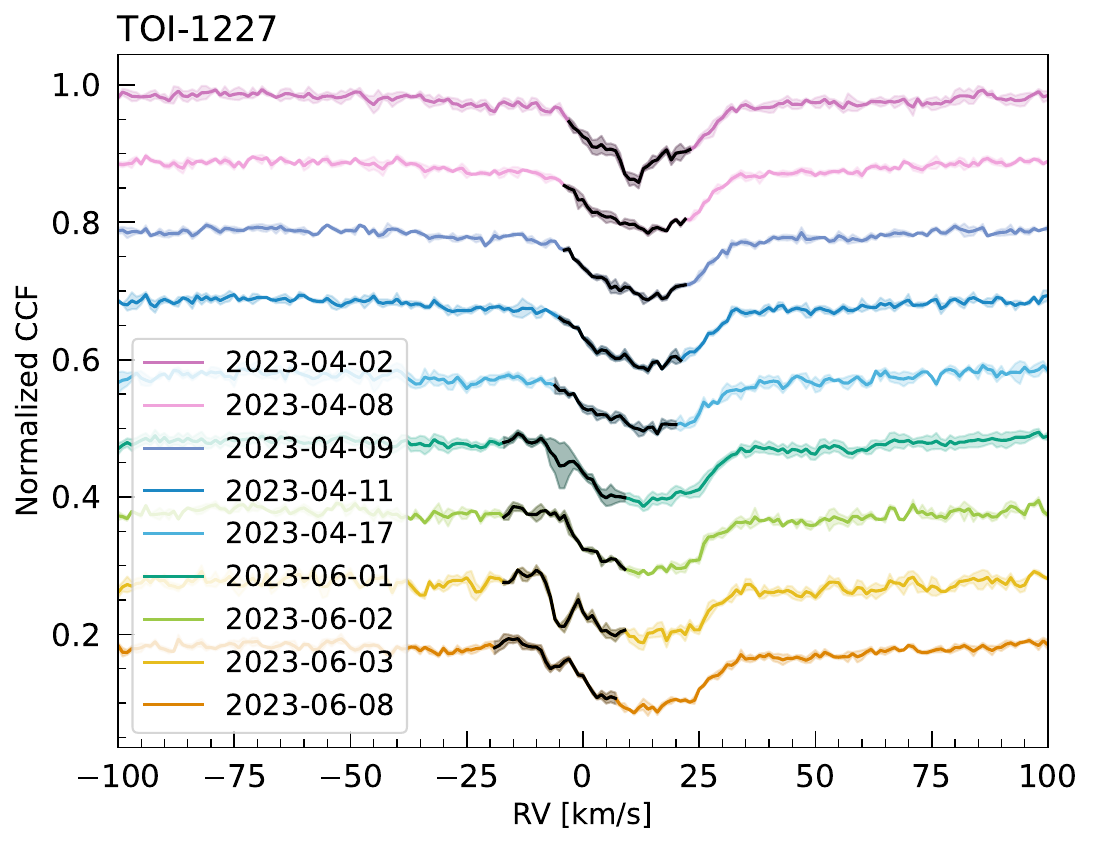}
    \caption{The normalized NIRPS CCFs of TOI-1227. The BERV affected region is indicated in black. The CCFs are broadened due to the fast stellar rotation.}
\end{figure}
\FloatBarrier

\subsection{TOI-2455}
TOI-2455 has a planetary radius of $R_{\rm pl} = 1.3\rjup$, which is relatively large for an M2V host star ($T_{\rm eff} =3553~{\rm K}$; \textit{TESS} Input Catalog). A significant RV variation would therefore be expected. However, TOI-2455 is a rapid rotator, broadening the CCFs to a FWHM of $\sim 40~{\rm km~s^{-1}}$. After one observation, which also coincided with a BERV crossing, the target was stopped.

\begin{figure}[H]
    \includegraphics[width=\linewidth]{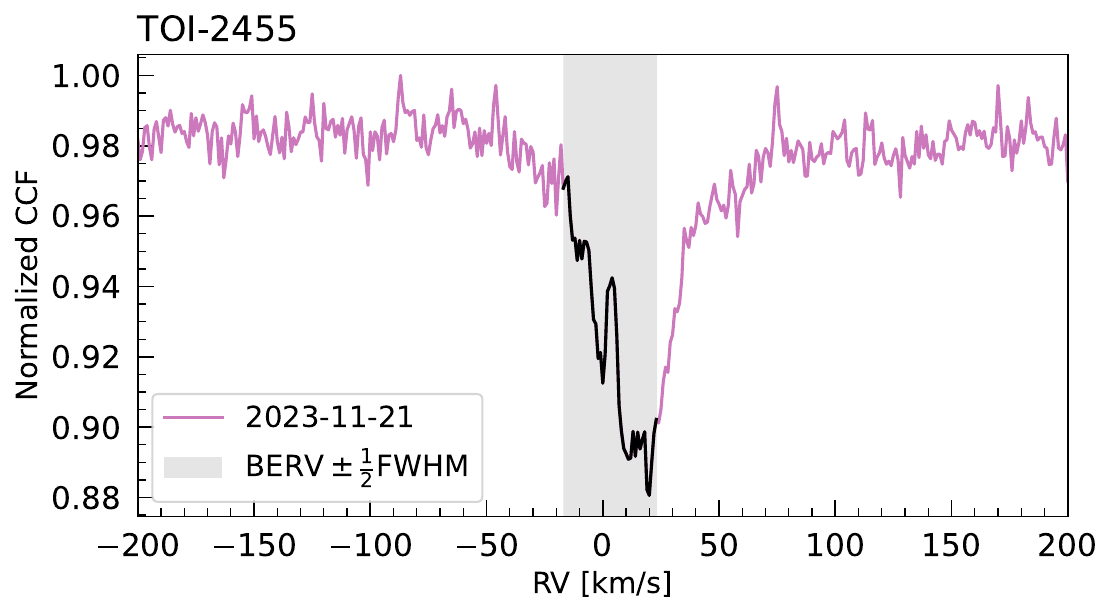}
    \caption{The normalized NIRPS CCFs of the fast rotating TOI-2455. The BERV affected region is highlighted in gray.}
\end{figure}
\FloatBarrier

\clearpage
\section{Giant stars}
\label{app:Giant_Stars}
The five TOIs that appear as evolved stars in the HR diagram in Fig.~\ref{fig:HRgaia} were part of the NIRPS-GTO program at its start (April 2023). Although transits are reported for these targets, the stellar radii of evolved stars are poorly constrained; consequently, the planetary radii inferred from the transit depths are likewise highly uncertain. We observed TOI-3463 once before recognizing it as a giant, and subsequently stopped its follow-up. For TOI-3209, the \textit{Gaia} DR3 General Stellar Parametrizer from Photometry (GSP-Phot) library reports a stellar radius of $36.3~R_\odot$, such that the $\sim 1~\%$ transit depth would correspond to a companion of $\sim 3.1~R_\odot$. For the other stars, no stellar radius is reported by \textit{Gaia}.

\FloatBarrier
\section{SED analysis}
\begin{figure}[H]
\centering
\begin{subfigure}[b]{0.975\linewidth}
    \includegraphics[width=\linewidth]{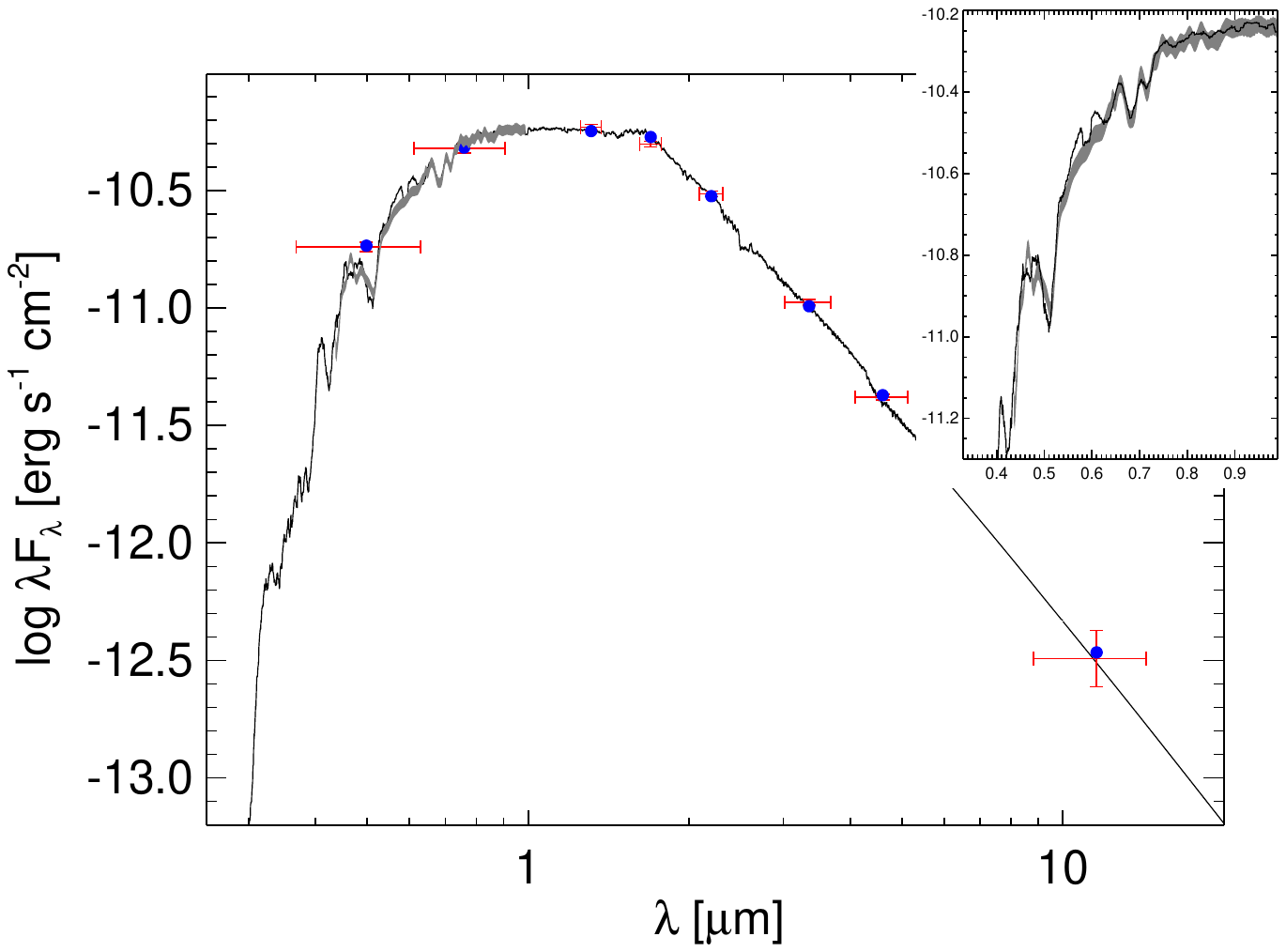}
    \caption{TOI-3288}
\end{subfigure}
\begin{subfigure}[b]{0.975\linewidth}
    \includegraphics[width=\linewidth]{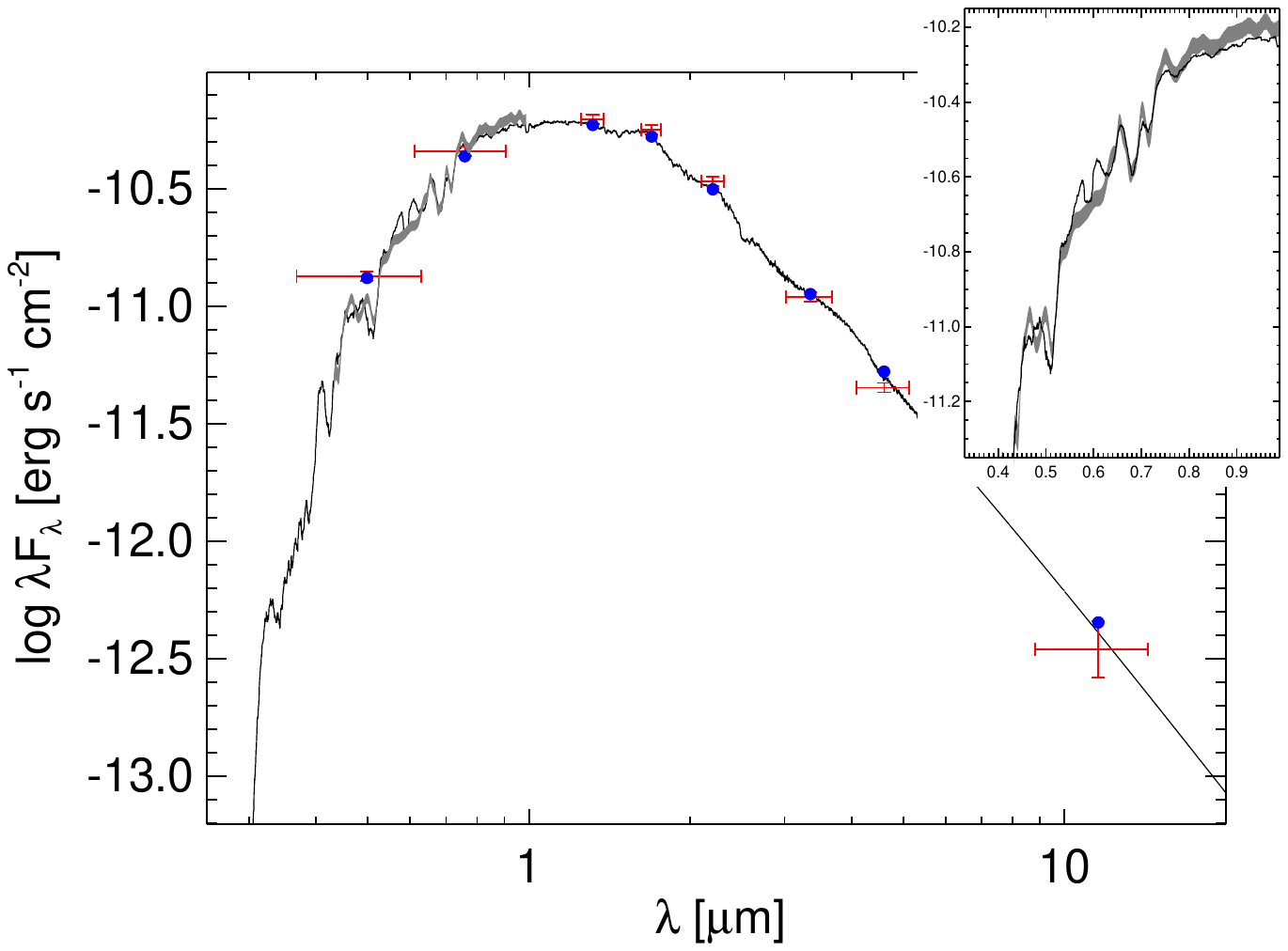}
    \caption{TOI-4666}
\end{subfigure}

\caption{The SEDs of the presented stars with giant companions. Red symbols represent the observed photometric measurements (as discussed in Section \ref{sec:SED_Analysis}), and the horizontal bars represent the effective width of the passband. Blue symbols are the model fluxes from the best-fit PHOENIX atmosphere model (black). The insets show the absolute flux-calibrated \textit{Gaia} spectrophotometry as a gray swathe overlaid on the model (black).}\label{fig:sed}
\end{figure}
\FloatBarrier
\newpage
\section{Guiding Frames}
\label{app:AO_Guiding}
\begin{figure}[H]
    \centering
    \begin{subfigure}[b]{.975\linewidth}
        \includegraphics[width=\linewidth]{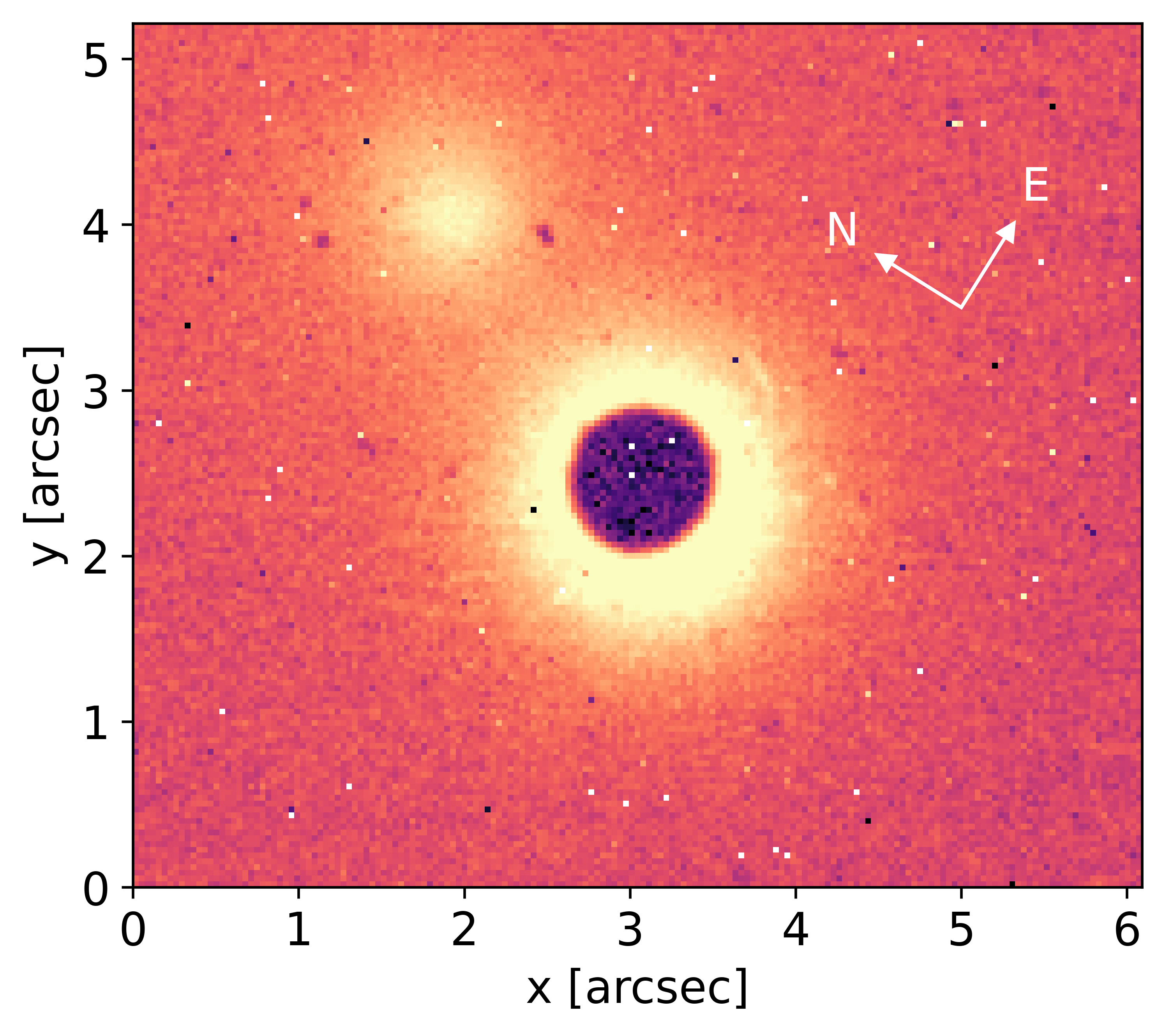}
    \end{subfigure}
    \hspace{.1\linewidth}
    \begin{subfigure}[b]{.975\linewidth}
        \includegraphics[width=\linewidth]{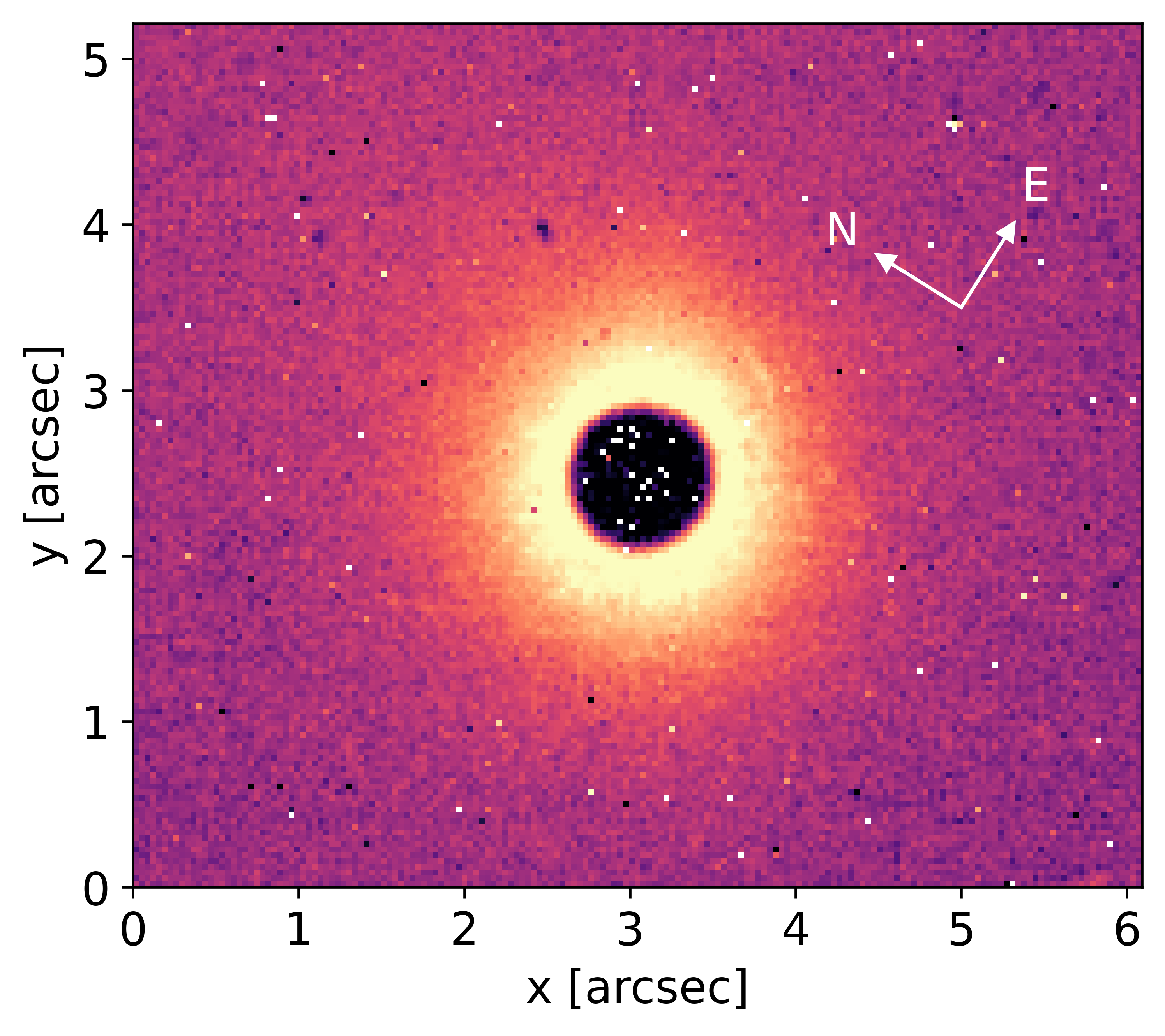}
    \end{subfigure}
    \caption{Integrated NIRPS AO guiding frames for TOI-3288 (top) and TOI-4666 (bottom). The stellar companion \textit{Gaia} DR3 6685431748040148992 is visible at a separation of $\sim2\arcsec$ from TOI-3288. No evidence of a stellar companion is seen for TOI-4666 (\textit{Gaia} DR3 $\rm{RUWE}=0.998$). The $0.9\arcsec$ fibre hole partially blocks the stellar light.}
    \label{fig:AO_Guiding}
\end{figure}

Integrated AO guiding frames are available per NIRPS observation. Figure~\ref{fig:AO_Guiding} shows the average frame over all observations for TOI-3288 and TOI-4666. A logarithmic scale is used to enhance the visibility of potential stellar companions. The stellar companion of TOI-3288 is visible at a separation of $\sim 2\arcsec$. This is the companion discussed in Sec.~\ref{sec:Binarity}, identified as \textit{Gaia} DR3 6685431748040148992.

\FloatBarrier
\clearpage
\section{Limb darkening and photometric instrumental parameters}
\label{app:limb_instrument_table}

\begin{table}[h!]
\caption{Fitted limb darkening parameters for the companions presented in this paper.}
\label{table:limb_darkening}
\centering
\begin{tabular}{l l l  c c c c c c}
\hline\hline
\multicolumn{2}{l}{}  & \multicolumn{3}{c}{TOI-3288 b} & \multicolumn{3}{c}{TOI-4666 b} \\ \hline
\multirow{2}{*}{\footnotesize{$\quad\circ\,\,$TESS}} & $q_\mathrm{1, TESS}$  & \multicolumn{3}{c}{$0.23_{-0.07}^{+0.11}$ } & \multicolumn{3}{c}{$0.10_{-0.04}^{+0.06}$ }\\
     & $q_\mathrm{2, TESS}$  & \multicolumn{3}{c}{$0.6\pm 0.2$ } & \multicolumn{3}{c}{$0.5\pm 0.3$ }\\
\multirow{2}{*}{\footnotesize{$\quad\circ\,\,$ExTrA}} & $q_\mathrm{1, ExTrA}$  & & & & \multicolumn{3}{c}{$0.07_{-0.02}^{+0.03}$ }\\
     & $q_\mathrm{2, ExTrA}$  & & & & \multicolumn{3}{c}{$0.6\pm 0.2$ }\\
\multirow{2}{*}{\footnotesize{$\quad\circ\,\,$LCO-CTIO ($g'$)}} & $q_{1, g'}$  & \multicolumn{3}{c}{$0.87_{-0.13}^{+0.09}$ } & & &\\
     & $q_{2, g'}$  & \multicolumn{3}{c}{$0.80_{-0.08}^{+0.11}$ } & & &\\
\multirow{2}{*}{\footnotesize{$\quad\circ\,\,$LCO-SAAO ($i'$)}} & $q_\mathrm{1, i'_1}$  & \multicolumn{3}{c}{$0.94_{-0.08}^{+0.04}$ } & & &\\
     & $q_\mathrm{2, i'_1}$  & \multicolumn{3}{c}{$0.25\pm 0.06$ } & & &\\
\multirow{2}{*}{\footnotesize{$\quad\circ\,\,$LCO-SSO ($i'$)}} & $q_{1, i'_2}$  & \multicolumn{3}{c}{$0.7\pm 0.2$ } & & &\\
     & $q_{2, i'_2}$  & \multicolumn{3}{c}{$0.6\pm 0.2$ } & & &\\ \hline
\end{tabular}
\end{table}

\begin{table}[h!]
\caption{Fitted photometric instrumental parameters for the companions presented in this paper.}
\label{table:photometric_instrument}
\begin{center}
\begin{tabular}{l l  c c c c c c}
\hline\hline
\multicolumn{1}{l}{} &  & \multicolumn{3}{c}{TOI-3288 b} & \multicolumn{3}{c}{TOI-4666 b} \\ \hline
$\gamma_\mathrm{TESS}$ & [$\times10^{-6}$]  & \multicolumn{3}{c}{$-245_{-36}^{+35}$ } & \multicolumn{3}{c}{$-196_{-65}^{+62}$ }\\
$\sigma_\mathrm{TESS}$ & [ppm]  & \multicolumn{3}{c}{$45_{-44}^{+359}$ } & \multicolumn{3}{c}{$16_{-15}^{+449}$ } \\
$\gamma_{g'}$ & [$\times10^{-3}$]  & \multicolumn{3}{c}{$105_{-25}^{+18}$ } & & &\\
$\sigma_{g'}$ & [ppm]  & \multicolumn{3}{c}{$10_{-10}^{+525}$ } & & &\\
$\theta_{0,g'}$ & [$\times10^{-3}$]  & \multicolumn{3}{c}{$103_{-20}^{+13}$ } & & &\\
$\gamma_{i'_1}$ & [$\times10^{-3}$]  & \multicolumn{3}{c}{$-7\pm 2$ } & & &\\
$\sigma_{i'_1}$ & [ppm]  & \multicolumn{3}{c}{$909_{-155}^{+68}$ } & & &\\
$\theta_{0,i'_1}$ & [$\times10^{-3}$]  & \multicolumn{3}{c}{$-7\pm 2$ } & & &\\
$\gamma_{i'_2}$ & [$\times10^{-3}$]  & \multicolumn{3}{c}{$-33\pm 20$ } & & &\\
$\sigma_{i'_2}$ & [ppm]  & \multicolumn{3}{c}{$12_{-12}^{+271}$ } & & &\\
$\theta_{0,i'_2}$ & [$\times10^{-3}$]  & \multicolumn{3}{c}{$-33_{-20}^{+19}$ } & & &\\ \hline
\end{tabular}
\end{center}
\textbf{Notes:} Where $\gamma$ is the offset relative flux, $\sigma$ the jitter, and $\theta_0$ the linear regressor. The filter subscripts correspond to the instruments specified for the limb darkening coefficients in Table \ref{table:limb_darkening}. ExTrA parameters can be found in Table \ref{table:photometric_instrument_ExTrA}.
\end{table}

\begin{table}
\caption{Fitted photometric instrumental parameters of ExTrA for TOI-4666.}
\label{table:photometric_instrument_ExTrA}
\begin{center}
\begin{tabular}{c r c c c}
\hline\hline
\multicolumn{1}{l}{} & & T1 & T2 & T3 \\ \hline
\multirow{10}{*}{\begin{tabular}{@{}c@{}}$\gamma$ \\ $[\times10^{-3}]$\end{tabular}} & \footnotesize{02-02-2022} & {$-9_{-22}^{+11}$} & {$-7_{-30}^{+28}$} & {$10_{-13}^{+39}$} \\
& \footnotesize{16-12-2022} & {$-1_{-14}^{+9}$} & {$0_{-9}^{+10}$} & {$0_{-7}^{+8}$} \\
& \footnotesize{21-09-2023} & {$-8_{-24}^{+9}$} & {$-33_{-43}^{+28}$} & {$-19_{-27}^{+22}$} \\
& \footnotesize{24-09-2023} & {$-0\pm2$} & {$0_{-6}^{+4}$} & {$38_{-21}^{+20}$} \\
& \footnotesize{29-10-2023} & {$1_{-6}^{+14}$} & {$-13_{-23}^{+14}$} & {$5_{-6}^{+21}$} \\
& \footnotesize{27-11-2023} & {$-77_{-30}^{+42}$} & {$-4_{-55}^{+5}$} & {$-7_{-27}^{+23}$} \\
& \footnotesize{30-11-2023} & {$-30_{-46}^{+30}$} & {$-16_{-24}^{+20}$} & {$-3_{-13}^{+6}$} \\
& \footnotesize{03-12-2023} & {$-1\pm3$} & {$3_{-5}^{+15}$} & {$2_{-6}^{+10}$} \\
& \footnotesize{01-01-2024} & {$-2_{-12}^{+6}$} & & {$1_{-3}^{+4}$} \\
& \footnotesize{04-01-2024} & {$-4_{-14}^{+15}$} & & {$-12_{-21}^{+12}$} \\ \hline
\multirow{10}{*}{\begin{tabular}{@{}c@{}}$\sigma$ \\ $[\mathrm{ppm}]$\end{tabular}} & \footnotesize{02-02-2022} & {$6_{-6}^{+135}$} & {$13_{-13}^{+196}$} & {$15_{-14}^{+166}$} \\
& \footnotesize{16-12-2022} & {$7_{-7}^{+102}$} & {$11_{-10}^{+149}$} & {$27_{-25}^{+224}$} \\
& \footnotesize{21-09-2023} & {$9_{-9}^{+136}$} & {$17_{-16}^{+217}$} & {$5_{-4}^{+87}$} \\
& \footnotesize{24-09-2023} & {$6_{-6}^{+115}$} & {$6_{-5}^{+83}$} & {$7_{-7}^{+111}$} \\
& \footnotesize{29-10-2023} & {$27_{-26}^{+265}$} & {$12_{-12}^{+197}$} & {$10_{-10}^{+158}$} \\
& \footnotesize{27-11-2023} & {$3_{-3}^{+40}$} & {$7_{-7}^{+102}$} & {$5_{-5}^{+69}$} \\
& \footnotesize{30-11-2023} & {$12_{-11}^{+168}$} & {$8_{-8}^{+116}$} & {$25_{-23}^{+224}$} \\
& \footnotesize{03-12-2023} & {$6_{-6}^{+102}$} & {$11_{-10}^{+160}$} & {$5_{-4}^{+83}$} \\
& \footnotesize{01-01-2024} & {$20_{-19}^{+225}$} & & {$9_{-9}^{+128}$} \\
& \footnotesize{04-01-2024} & {$14_{-13}^{+174}$} & & {$16_{-15}^{+189}$} \\ \hline
\multirow{10}{*}{\begin{tabular}{@{}c@{}}$\sigma_{\rm GP}$ \\ $[\mathrm{ppm}]$\end{tabular}} & \footnotesize{02-02-2022} &   {$0.02_{-0.01}^{+0.04}$ } &   {$0.06_{-0.03}^{+0.11}$ } &   {$0.02_{-0.01}^{+0.05}$ } \\
& \footnotesize{16-12-2022} &   {$0.02_{-0.01}^{+0.06}$ } &   {$0.013_{-0.005}^{+0.016}$ } &   {$0.009_{-0.005}^{+0.026}$ } \\
& \footnotesize{21-09-2023} &   {$0.02_{-0.01}^{+0.04}$ } &   {$0.04_{-0.03}^{+0.08}$ } &   {$0.07_{-0.05}^{+0.23}$ } \\
& \footnotesize{24-09-2023} &   {$0.005_{-0.001}^{+0.003}$ } &   {$0.007_{-0.003}^{+0.008}$ } &   {$0.11_{-0.07}^{+0.41}$ } \\
& \footnotesize{29-10-2023} &   {$0.009_{-0.006}^{+0.030}$ } &   {$0.03_{-0.02}^{+0.09}$ } &   {$0.010_{-0.009}^{+0.038}$ } \\
& \footnotesize{27-11-2023} &   {$0.10_{-0.06}^{+0.17}$ } &   {$0.008_{-0.005}^{+0.116}$ } &   {$0.05_{-0.03}^{+0.10}$ } \\
& \footnotesize{30-11-2023} &   {$0.04_{-0.03}^{+0.13}$ } &   {$0.05_{-0.03}^{+0.12}$ } &   {$0.011_{-0.006}^{+0.018}$ } \\
& \footnotesize{03-12-2023} &   {$0.007_{-0.002}^{+0.003}$ } &   {$0.012_{-0.009}^{+0.044}$ } &   {$0.010_{-0.006}^{+0.019}$ } \\
& \footnotesize{01-01-2024} &   {$0.011_{-0.006}^{+0.018}$ } & &   {$0.004_{-0.002}^{+0.007}$ } \\
& \footnotesize{04-01-2024} &   {$0.03_{-0.02}^{+0.07}$ } & &   {$0.03_{-0.02}^{+0.07}$ } \\ \hline
\multirow{10}{*}{\begin{tabular}{@{}c@{}}$\rho_{\rm GP}$ \\ $[\mathrm{d}]$\end{tabular}} & \footnotesize{02-02-2022} &   {$0.15_{-0.08}^{+0.23}$ } &   {$0.5_{-0.3}^{+0.7}$ } &   {$0.2_{-0.1}^{+0.4}$ } \\
& \footnotesize{16-12-2022} &   {$0.8_{-0.5}^{+1.9}$ } &   {$0.07_{-0.03}^{+0.06}$ } &   {$0.10_{-0.06}^{+0.38}$ } \\
& \footnotesize{21-09-2023} &   {$0.2_{-0.1}^{+0.3}$ } &   {$0.4_{-0.2}^{+0.5}$ } &   {$2_{-1}^{+5}$ } \\
& \footnotesize{24-09-2023} &   {$0.015_{-0.007}^{+0.027}$ } &   {$0.07_{-0.03}^{+0.09}$ } &   {$4_{-3}^{+14}$ } \\
& \footnotesize{29-10-2023} &   {$0.10_{-0.07}^{+0.28}$ } &   {$1.1_{-0.9}^{+4.2}$ } &   {$25_{-23}^{+237}$ } \\
& \footnotesize{27-11-2023} &   {$2_{-2}^{+9}$ } &   {$0.06_{-0.05}^{+8.89}$ } &   {$0.6_{-0.3}^{+0.9}$ } \\
& \footnotesize{30-11-2023} &   {$0.5_{-0.4}^{+1.4}$ } &   {$0.8_{-0.5}^{+1.4}$ } &   {$0.16_{-0.08}^{+0.22}$ } \\
& \footnotesize{03-12-2023} &   {$0.018_{-0.005}^{+0.010}$ } &   {$0.4_{-0.4}^{+2.3}$ } &   {$0.09_{-0.06}^{+0.17}$ } \\
& \footnotesize{01-01-2024} &   {$0.11_{-0.07}^{+0.16}$ } & &   {$0.09_{-0.05}^{+0.13}$ } \\
& \footnotesize{04-01-2024} &   {$0.6_{-0.3}^{+0.9}$ } & &   {$0.7_{-0.6}^{+2.1}$ } \\ \hline
\end{tabular}
\end{center}
\textbf{Notes:} Where $\gamma$ is the offset relative flux, $\sigma$ the jitter, $\sigma_{\rm GP}$ the amplitude of the GP, and $\rho_{\rm GP}$ the Matern time-scale.
\end{table}

\clearpage
\section{Joint modeling priors}
\label{app:priors}
\begin{table}[h!]
            \caption{Priors for the joint modeling of photometric and radial velocity data.}
            \label{table:priors}
            \begin{center}
    \resizebox{\linewidth}{!}{
            \begin{tabular}{l l  c c }
            \hline\hline
            \multicolumn{2}{l}{Parameter}  & TOI-3288 & TOI-4666 \\ \hline
            $P$ & [days]  & $\mathcal{U}$ (0.9, 1.9) & $\mathcal{U}$ (2.4, 3.4) \\
            $T_0$ & [BJD]  & $\mathcal{N}$ (2459057.7, 1.0) & $\mathcal{N}$ (2459168.5, 1.0) \\
            $R_\mathrm{pl}/R_{\star}$ &  & $\mathcal{U}$ (0, 1) & $\mathcal{U}$ (0, 1)\\
            $b$ &  & $\mathcal{U}$ (0, 1) & $\mathcal{U}$ (0, 1)\\
            $\rho_{\star}$ & [kg m$^{-3}$]  & $\mathcal{N}$ (3009, 532) & $\mathcal{N}$ (4030, 662) \\
            $q_\mathrm{1, TESS}$ & & $\mathcal{U}$ (0, 1)& $\mathcal{U}$ (0, 1)\\
            $q_\mathrm{2, TESS}$ & & $\mathcal{U}$ (0, 1)& $\mathcal{U}$ (0, 1)\\
            $q_\mathrm{1, phot}$ &  & $\mathcal{U}$ (0, 1) & $\mathcal{U}$ (0, 1)\\
            $q_\mathrm{2, phot}$ &  & $\mathcal{U}$ (0, 1) & $\mathcal{U}$ (0, 1)\\
            $D$ &  & $\mathcal{U}$ (0.1, 1) &  1\\
            $e$ &  &  0 &  0\\
            $\omega$ & [$^{\circ}$]  &  90 &  90\\
            $K$ & [km s$^{-1}$]  & $\mathcal{U}$ (0, 100) & $\mathcal{U}$ (0, 100)\\
            $\gamma_\mathrm{TESS}$ & & $\mathcal{N}$ (0, 0.1)& $\mathcal{N}$ (0, 0.1)\\
            $\sigma_\mathrm{TESS}$ & [ppm] & $\log\mathcal{U}$ (0.1, 1000)& $\log\mathcal{U}$ (0.1, 1000)\\
            $\gamma_\mathrm{phot}$ & & $\mathcal{N}$ (0, 0.1) & $\mathcal{N}$ (0, 0.1)  \\
            $\sigma_\mathrm{phot}$ & [ppm] & $\log\mathcal{U}$ (0.1, 1000) & $\log\mathcal{U}$ (0.1, 1000)  \\
            $\theta_{0,\mathrm{phot}}$ & & $\mathcal{U}$ (-100, 100) & \\
            $\sigma_{\rm GP}$ & [ppm] & & $\log \mathcal{U}(10^{-6}, 10^6)$\\
            $\rho_{\rm GP}$ & [d] & & $\mathcal{U}(10^{-3}, 1000)$\\
            $\gamma_\mathrm{RV}$ & [km s$^{-1}$] & $\mathcal{U}$ (-100, 100)& $\mathcal{U}$ (-100, 100)\\
            $\sigma_\mathrm{RV}$ & [km s$^{-1}$] & $\log\mathcal{U}$ ($10^{-6}$, 0.1)& $\log\mathcal{U}$ ($10^{-6}$, 0.1)\\ 
            \hline
            \end{tabular}}
            \end{center}
            \textbf{Notes:} For all ground-based photometric datasets, the limb-darkening coefficients ($q_\mathrm{1,phot}$ and $q_\mathrm{2,phot}$), relative flux offset ($\gamma_\mathrm{phot}$), jitter term ($\sigma_\mathrm{phot}$), and linear regressor ($\theta_\mathrm{0,phot}$) are fitted independently, but each is assigned the same prior distribution. The informative priors of $P$, $T_0$, and $\rho_{\star}$ follow from the {\tt EXOFOP-TESS} website\footref{foot:EXOFOP}.
            
        \end{table}

\FloatBarrier
\section{Radial velocities}
\label{app:RVs}
\begin{table}[H]
    \centering
    \caption{RVs of TOI-3288. The complete table
is available at the CDS.}
    \label{tab:TOI-3288_RVs}
    \begin{tabular}{c c c r}\hline\hline
    Time & RV & RV error & Instrument-Mode \\
    ${\rm [rBJD~TDB]}$ & [${\rm m~s^{-1}}$] & [${\rm m~s^{-1}}$] \\ \hline  
60427.8702 & 30067.4 & 93.6 & NIRPS-HE\\ 
60427.8801 & 30329.7 & 103.5 & NIRPS-HE\\
60427.8874 & 29850.6 & 90.0& NIRPS-HE\\  
60427.8801 & 30095.3 & 79.1 & HARPS-EGGS\\ 
$\dots$ & $\dots$ & $\dots$ & \multicolumn{1}{c}{$\dots$}\\
 \hline
    \end{tabular}
\end{table}
\begin{table}[H]
    \centering
    \caption{RVs of TOI-4666. The complete table is available at the CDS.}
    \label{tab:TOI-4666_RVs}
    \begin{tabular}{c c c r}\hline\hline
    Time & RV & RV error & Instrument-Mode \\
    ${\rm [rBJD~TDB]}$ & [${\rm m~s^{-1}}$] & [${\rm m~s^{-1}}$] \\ \hline  
60149.8994 & 7444.8 & 58.9 & NIRPS-HE\\ 
60149.9096 & 7049.7 & 91.7 & NIRPS-HE\\
60149.9189 & 7414.4 & 68.9 & NIRPS-HE\\ 
60149.9096 & 8025.7 & 98.1 & HARPS-EGGS\\
$\dots$ & $\dots$ & $\dots$ & \multicolumn{1}{c}{$\dots$ }\\
 \hline
    \end{tabular}
\end{table}
\textcolor{white}{~}\vfill
\newpage
\section{Photometry}
\label{app:photometry}
\begin{table}[H]
    \centering
    \caption{TESS photometry of TOI-3288 extracted as described in Section~\ref{sec:TESSphotometry}. The complete table is available at the CDS.}
    \label{tab:TOI-3288_phot}
    \begin{tabular}{c c c c}\hline\hline
    Time & Flux & Flux error & Sector\\
    ${\rm [rBJD~TDB]}$ & & & \\ \hline  
    58657.7243 & 0.9922 & 0.0030 & 13\\
    58657.7451 & 0.9898 & 0.0030 & 13\\
    $\dots$ & $\dots$ & $\dots$ & $\dots$\\\hline
\end{tabular}
\end{table}
\begin{table}[H]
    \centering
    \caption{TESS photometry of TOI-4666 extracted as described in Section~\ref{sec:TESSphotometry}. The complete table is available at the CDS.}
    \label{tab:TOI-4666_phot}
    \begin{tabular}{c c c c}\hline\hline
    Time & Flux & Flux error & Sector\\
    ${\rm [rBJD~TDB]}$ & &  &  \\ \hline  
59144.5205 & 0.9945 & 0.0048 & 31\\
59144.5275 & 0.9920 & 0.0048 & 31\\
$\dots$ & $\dots$ & $\dots$ & $\dots$\\ \hline 
\end{tabular}
\end{table}

\FloatBarrier
\onecolumn
\section{Corner plots}
\label{app:Corner}
\begin{figure}[H]
    \includegraphics[width=\linewidth]{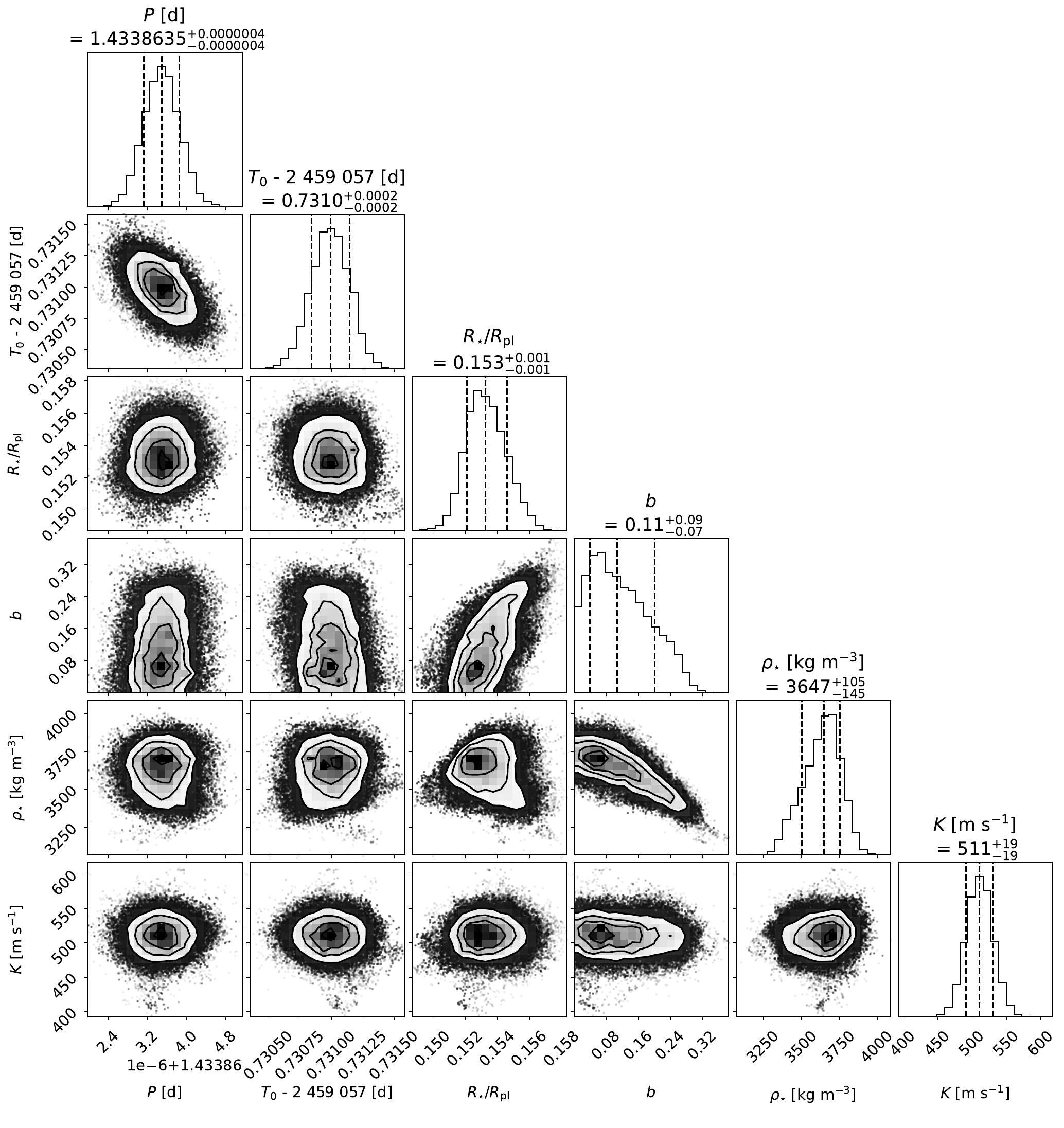}
    \caption{The corner plot for the TOI-3288 \texttt{Juliet} results.}
\end{figure}
\newpage
\begin{figure}[H]
    \includegraphics[width=\linewidth]{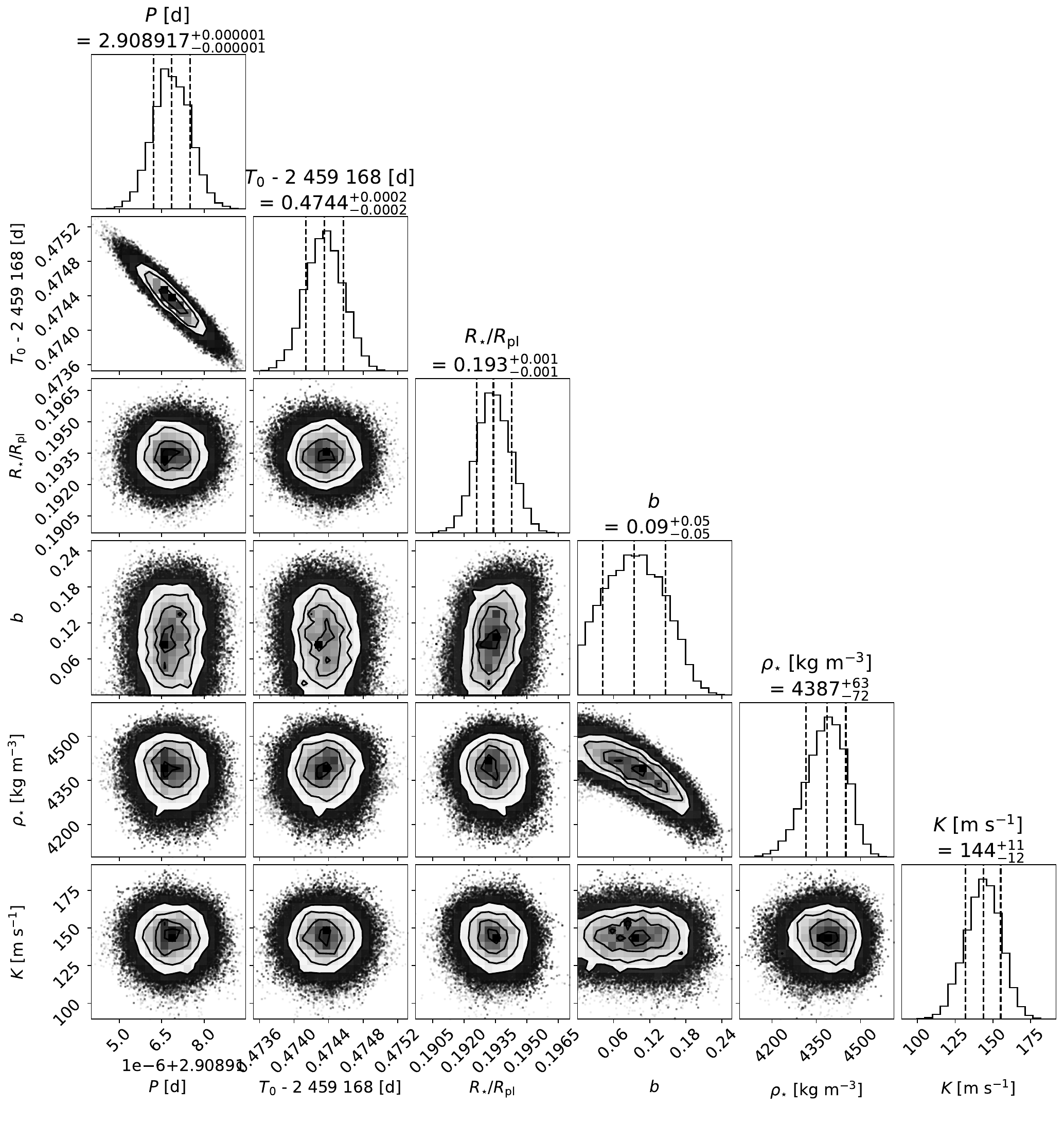}
    \caption{The corner plot for the TOI-4666 \texttt{Juliet} results.}
\end{figure}

\twocolumn
\clearpage
\section{\textit{TESS}-cont generated heatmaps}
\label{app:TESS-cont}
\begin{figure}[H]
    \includegraphics[width=\linewidth]{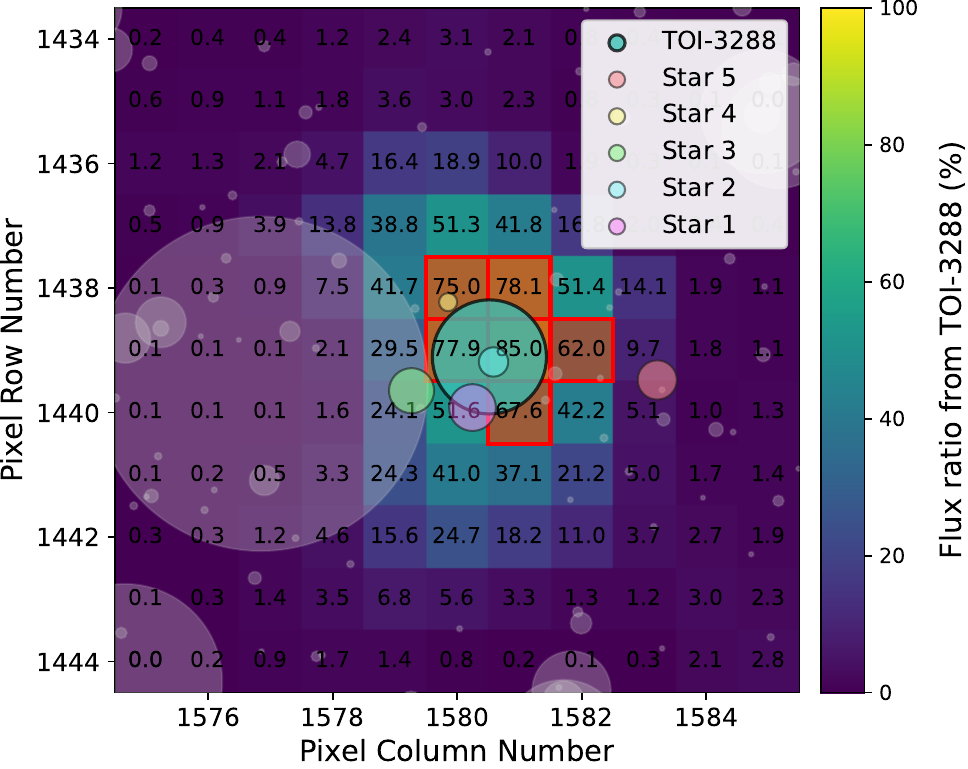}
    \caption{Sector 13 for TOI-3288. The flux percentage of TOI-3288 is indicated per pixel as an overlay of the \textit{TESS} FFI. The red box indicates the aperture used to extract the lightcurve, this aperture is $23~\%$ contaminated by nearby stars.}
\end{figure}

\begin{figure}[H]    \includegraphics[width=\linewidth]{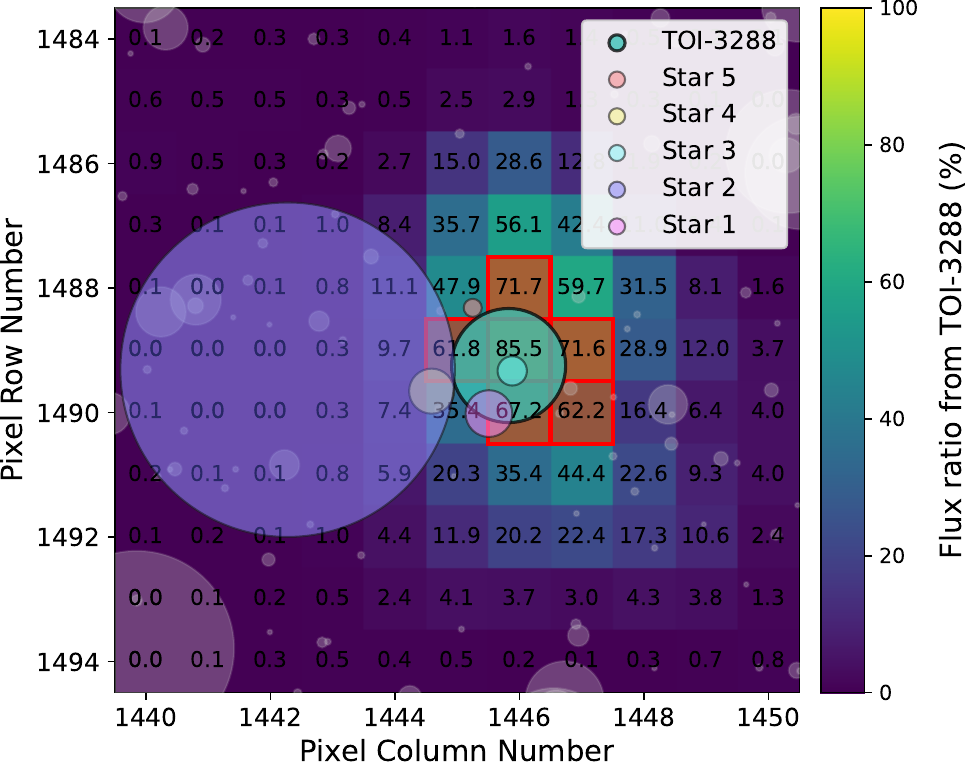}
    \caption{Sector 27 for TOI-3288. The flux percentage of TOI-3288 is indicated per pixel as an overlay of the \textit{TESS} FFI. The red box indicates the aperture used to extract the lightcurve, this aperture is $27~\%$ contaminated by nearby stars.}
\end{figure}

\newpage
\begin{figure}[H]    \includegraphics[width=\linewidth]{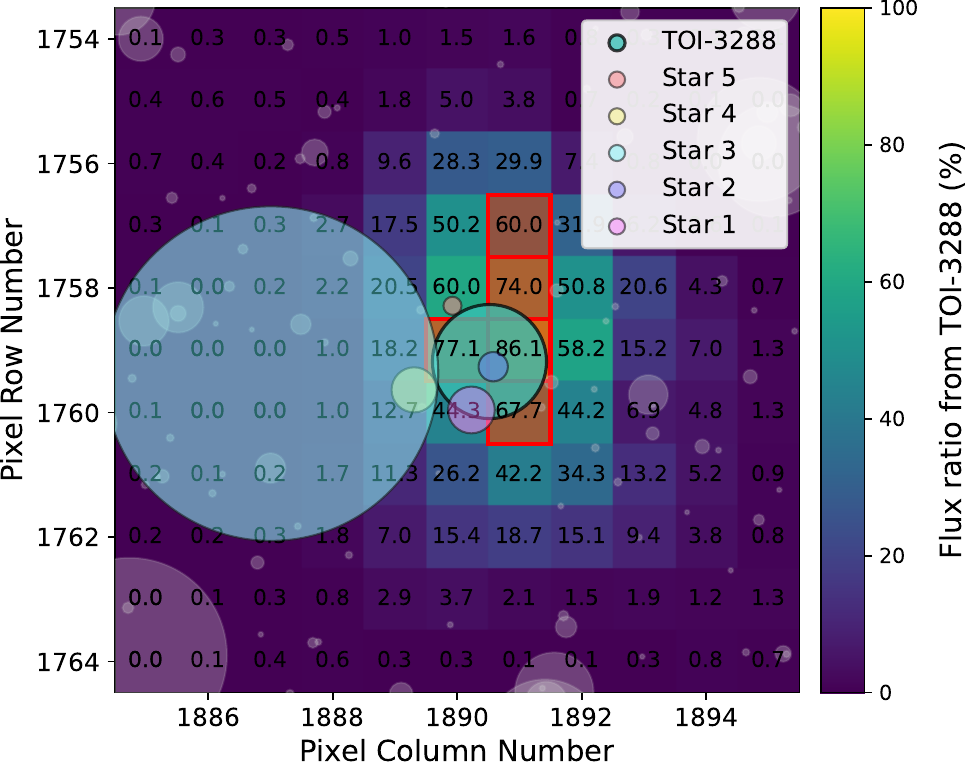}
    \caption{Sector 67 for TOI-3288. The flux percentage of TOI-3288 is indicated per pixel as an overlay of the \textit{TESS} FFI. The red box indicates the aperture used to extract the lightcurve, this aperture is $22~\%$ contaminated by nearby stars.}
\end{figure}

\begin{figure}[H]    \includegraphics[width=\linewidth]{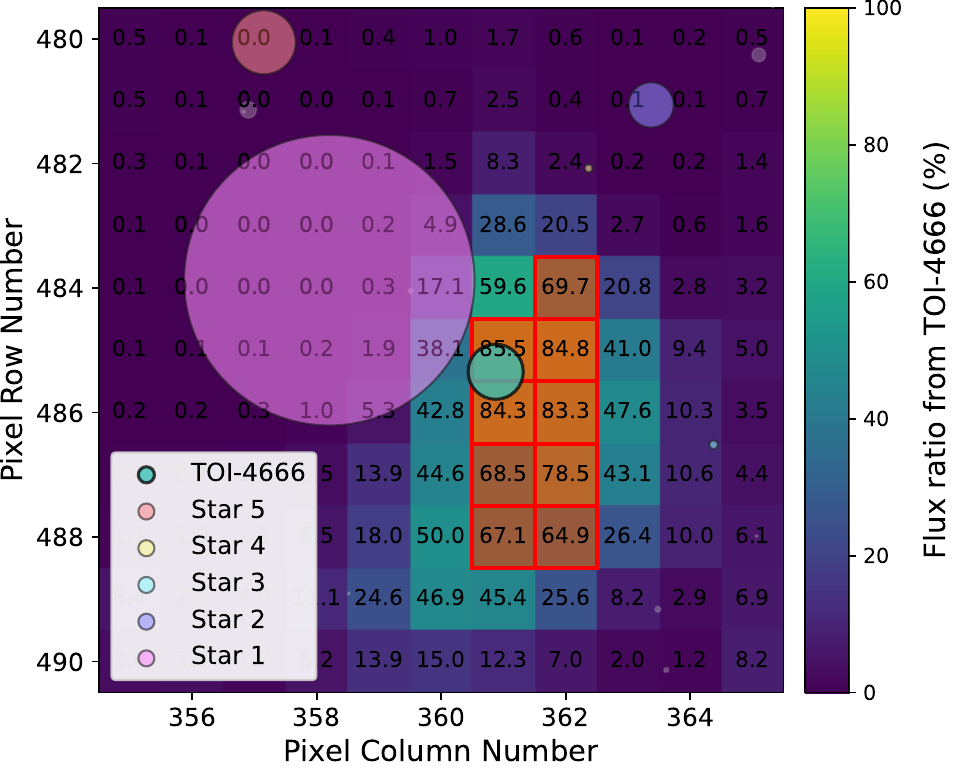}
    \caption{Sector 31 for TOI-4666. The flux percentage of TOI-4666 is indicated per pixel as an overlay of the \textit{TESS} FFI. The red box indicates the aperture used to extract the lightcurve, this aperture is $18~\%$ contaminated by nearby stars.}
\end{figure}

\clearpage
\onecolumn
\section{ASAS-SN periodograms}
\begin{figure}[H]
    \includegraphics[width=\linewidth]{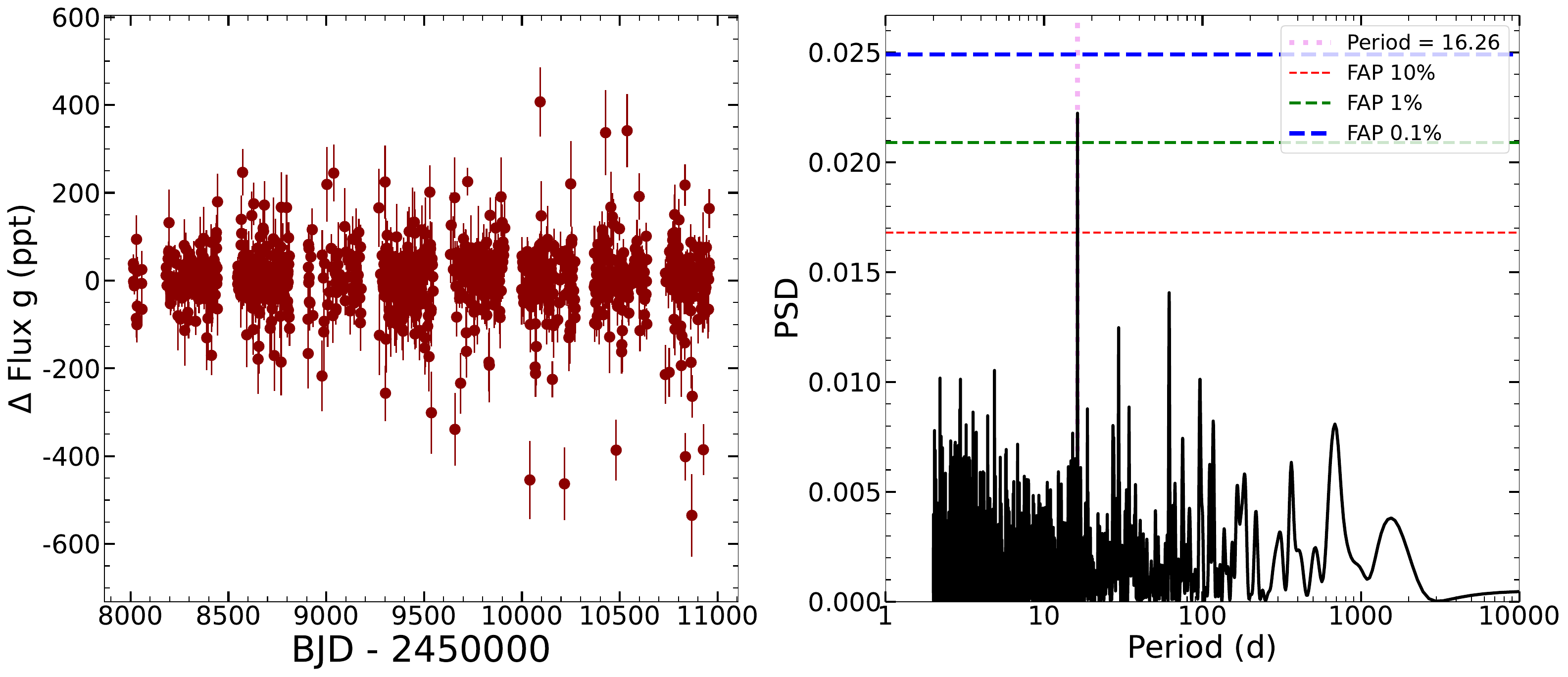}
    \caption{The $g$-band ASAS-SN data of TOI-3288 and the corresponding Lomb–Scargle periodogram.}
    \label{fig:asassn1}
\end{figure}
\begin{figure}[H]
    \includegraphics[width=\linewidth]{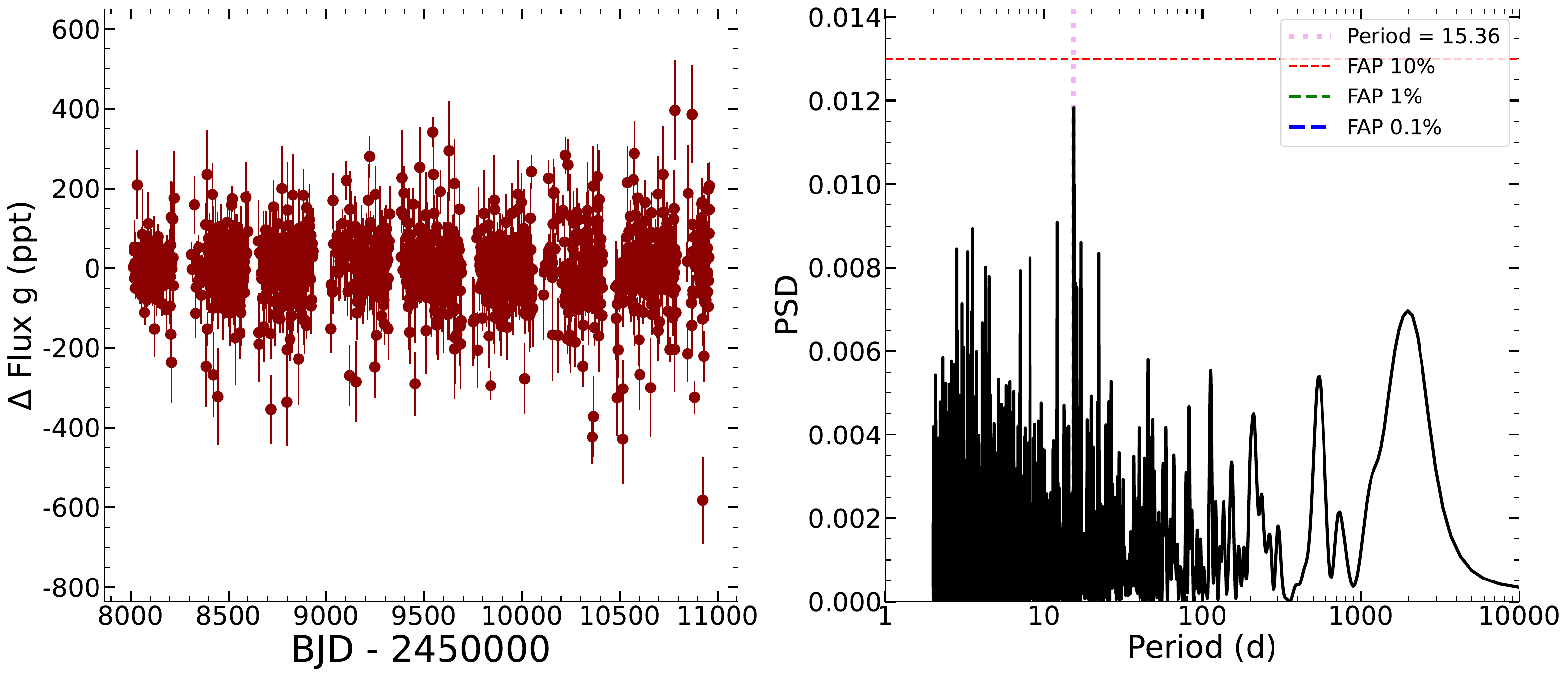}
    \caption{The $g$-band ASAS-SN data of TOI-4666 and the corresponding Lomb–Scargle periodogram.}
    \label{fig:asassn2}
\end{figure}

\end{document}